# TITLE

Liquid Flow Reversibly Creates a Macroscopic Surface Charge Gradient

## Authors

Patrick Ober[1], Willem Q. Boon[2], Marjolein Dijkstra[3], Ellen H. G. Backus[1,4], René van Roij[2*] & Mischa Bonn[1*]
[1]Department of Molecular Spectroscopy, Max Planck Institute for Polymer Research, Ackermannweg 10, 55128 Mainz, Germany. [2]Institute for Theoretical Physics, Utrecht University, Princetonplein 5, 3584 CC Utrecht, Netherlands. [3]Soft Condensed Matter, Debye Institute for Nanomaterials Science, Utrecht University, Princetonplein 5, 3584 CC Utrecht, Netherlands. [4]Department of Physical Chemistry, University of Vienna, Waehringer Strasse 42, 1090 Vienna, Austria.
These authors contributed equally: Patrick Ober, Willem Q. Boon
*email: r.vanroij@uu.nl, bonn@mpip-mainz.mpg.de

## Abstract

The charging and dissolution of mineral surfaces in contact with flowing liquids are ubiquitous in nature, as most minerals in water spontaneously acquire charge and dissolve. Mineral dissolution has been studied extensively under equilibrium conditions, even though non-equilibrium phenomena are pervasive and substantially affect the mineral-water interface. Here we demonstrate using interface-specific spectroscopy that liquid flow along a calcium fluoride surface creates a reversible spatial charge gradient, with decreasing surface charge downstream of the flow. The surface charge gradient can be quantitatively accounted for by a reaction-diffusion-advection model, which reveals that the charge gradient results from a delicate interplay between diffusion, advection, dissolution, and desorption/adsorption. The underlying mechanism is expected to be valid for a wide variety of systems, including groundwater flows in nature and microfluidic systems.

# Introduction

Reactions at the interface between a charged solid surface and a flowing fluid play a key role on macroscopic scales in geochemical cycles[1-4], as well as on microscopic scales in micro- and nanofluidic systems[5]. They are also central in technological applications in areas as diverse as froth flotation[6], electrophoresis[7], water desalination[8], soil remediation[9], and even dentistry[10]. Only a few years ago, LIS et al.[4] presented the first experimental evidence that the surface potential of mineral surfaces (silica and calcium fluoride) in contact with water changes substantially when liquid flow is applied. This observation implies that fluid flow can directly affect a chemical equilibrium. For the mineral fluorite ($CaF_2$) under acidic conditions, the change of the surface charge upon flow was argued to be due to the dilution of reactive ions partaking in the surface charging reaction[4]. This dilution is caused by the concentration difference between the fresh solution from the reservoir and that in the flow channel. The dilution changes the charging equilibrium and increases the surface charge, which manifests itself by an increase of the v-SFG response. A similar explanation has also recently been given by XI et al.[11]. Even though the flow-induced disturbance of the charging equilibrium qualitatively explains several experimental features, there are also inconsistencies. For instance, the question remains how a concentration difference between the reservoir and channel is generated. The surface charging reaction itself cannot supply the excess of ions in the channel over the bulk concentration, as the number of reactive ions on the surface is too small to support a steady flux of ions, especially over many flow cycles. Therefore, a fully self-consistent quantitative model is missing. In fact, several different hypotheses have been put forward for the change of the v-SFG signal upon flow, for instance, a surface conduction model by WERKHOVEN et al.[12], which would lead to a surface charge gradient and a 1D toy model by LIAN et al.[11] that considers net dissolution as the driving force. Moreover, SCHAEFER et al.[13] concluded that a change in surface potential could be explained by a flow-induced change of the concentration of ions that screen the surface charge.

A flow-dependent surface charge not only paves the way for novel electrokinetic effects[14], but it also has important consequences for the interpretation of zeta potential measurements. However, the mechanism by which the flow alters the surface charge is poorly understood as of yet[15].

One needs an interface-specific method to obtain a fundamental understanding of the events occurring at interfaces. At solid-water interfaces, vibrational Sum Frequency Generation (v-SFG) is a well-established method that is interface-specific and provides molecular information[2,4,13,16-27]. The interface-specificity of v-SFG results from the selection rule that inversion symmetry must be broken for a v-SFG signal to be generated. If the solid-liquid interface is charged, the resulting electric field aligns the dipole moments of the interfacial water molecules. Additionally, the electric field also polarizes the water molecules some nanometers into the solution. Since both phenomena break the symmetry, the strength of the observed v-SFG signal increases with the average polarization and orientation of the water molecules. Therefore, the v-SFG response is a measure of the electric field at the surface (see Supplementary Note 1 for details)[2,4]. By Gauss' law, this field is connected to the surface charge and is directly related to the surface potential.

In this work, we investigate the flow-dependent surface potential by measuring v-SFG spectra at several positions in a channel with a length of 2.48 cm and a radius of 0.24 cm. The local changes in the surface potential upon flow are tracked through the intensity of the OH-stretch band in v-SFG spectra, which reports on the alignment and polarization of interfacial water molecules and, therefore, the surface charge. Using this approach, we show that the surface charge not only depends on the flow rate but also on the position in the channel, along the flow direction. We explain these observations both qualitatively and quantitatively in terms of coupled reaction-diffusion-advection processes that give rise to flow-dependent heterogeneous ion concentrations, causing the surface charge of $CaF_2$ to vary spatially along the flow direction in the channel. Our theoretical model reveals that the macroscopic lateral gradient of the surface charge density (and hence of the v-SFG response) along the mineral can be traced to a gradient in the interfacial fluoride concentration. This fluoride gradient also causes a gradient in the dissolution rate along the mineral surface. Such a gradient of the net dissolution rate in confined media is argued to be important for understanding geochemical processes, for instance, in groundwater flows[28,29]. Additionally, our approach of combining a full numerical model with position-resolved v-SFG experiments allows determining whether the surface is charged by desorption or adsorption when the dissolution mechanism is known.

## Results

**Flow-induced Change in Surface Charge**

Similar to LIS et al.[4] and XI et al.[26], we conducted flow experiments at the $CaF_2$-water interface under acidic conditions, first at a single position in the channel. Fig. 1a illustrates our experiment. Visible (Vis) and infrared (IR) laser pulses overlap in space and time at the $CaF_2$-water interface, such that a nonlinear optical v-SFG process generates a response at a frequency equal to the sum of the two incident frequencies. Due to spectroscopic selection rules, this process is forbidden in centrosymmetric media such as bulk water. In contrast to the bulk, centrosymmetry is broken at the interface, which renders the v-SFG signal interface-specific[16,17]. Moreover, the resonance of the IR pulse with the water stretching mode at around 3300 cm$^{-1}$ enhances the overall process and provides additional molecular specificity[2,4,18-25].

Our measurements are performed in total internal reflection geometry to increase the signal and reduce the acquisition time, allowing time resolution on a ~second time scale. Fig. 1b shows two v-SFG spectra of the $CaF_2$-water interface for a 1 mM NaCl and 1 mM HCl (so pH 3) solution, once under flow-off and once under flow-on conditions. The measured signal is plotted *vs.* the wavenumber of the IR pulse being in resonance with the OH stretch vibration of water molecules. The flow-on (red) spectrum in Fig. 1b is ~40% higher in intensity compared to the flow-off (blue) spectrum in Fig. 1b. This increase can be correlated to an increase in surface charge (see Supplementary Note 1 for details). Thus, we reproduced that flow increases the surface charge at the $CaF_2$-water interface under acidic conditions[4,26]. The underlying mechanism for the flow-induced increase in the surface charge will be considered in the discussion section. When we calculate the intensity of the two spectra of Fig. 1b by integrating over the spectral area, we can track their time-dependence upon switching on and off the flow with a resolution of seconds, shown in Fig. 1c, where each black circle stems from one spectrum. We note that the v-SFG response is constant before the liquid flow is switched on, which therefore represents a steady state. Upon applying flow, another steady state with an increased v-SFG response is reached within about 5 minutes. When the flow is turned off again, the initial steady state is restored but only after about 10 minutes. As explained above, the v-SFG response can be used as a measure for the surface charge. Thus, the flow-induced change in the v-SFG signal can be correlated to a reversible change in surface charge due to flow. The observation of reversible changes in the v-SFG response and thus in surface charge due to liquid flow is consistent with the studies of LIS et al.[4] and XI et al.[26].

**Observation of Flow-Induced Surface Charge Gradient**

In order to gain an understanding of the interfacial events along the mineral surface upon flow, we compare the changes in v-SFG response at different positions along the flow channel. Another approach is to compare the change in v-SFG signal upon flow at one position, which is not the center of the flow cell, and reversing the flow direction, i.e. interchanging the role of the inlet and outlet as illustrated in Fig. 2a. Following the second approach, time traces of the integrated v-SFG spectra at three different positions with flow on/off cycles in clockwise and counter-clockwise flow directions are shown in Fig. 2b. First of all, it can be seen that the v-SFG response increases upon flow at every position. Additionally, we observe that at the center of the flow channel (orange spot and trace in Fig. 2b) essentially identical changes of the v-SFG intensity for both flow directions are observed, which is fully consistent with the study of LIS et al.[4]. Here we do not obtain additional spatial information as reversing the flow direction interchanges the center with itself due to the flow cell's symmetry. In contrast, at the two opposite sites at equal distance from the center (8 mm), the reversible changes upon flow depend on the flow direction. At a clockwise flow direction, the increase in the v-SFG response at the red position in Fig. 2 is substantially and reproducibly higher than at a counter-clockwise flow direction, while the opposite is the case at the other side of the flow cell (blue spot and blue trace). We also note that the magnitude of the increase in v-SFG response in the center is approximately in between that at the inlet and outlet. Thus, we conclude that the increase in the v-SFG response upon flow decreases in the flow direction, such that our measurements show that along the mineral there is a flow-induced surface charge gradient in the flow direction.

To further investigate the proposed gradient along the flow channel, we quantitatively compare the flow-induced changes at different positions along the centimeter-sized flow channel for the same flow direction. To do so, we focus on the relative increase of the v–SFG response under flow-on conditions (compared to flow-off) as a function of the distance from the center of the channel. Additionally, we study the influence of the flow rate. The relative increase in v-SFG intensity $\left(\frac{I_{\text{On}}-I_{\text{Off}}}{I_{\text{Off}}}\right)$ accounts for slight changes in the alignment that necessarily occur when changing the position of the flow cell. The position-dependent relative increase in intensity is determined by averaging the intensity in steady states, as indicated by the marked regimes in Fig. 1c. The obtained data are shown in Fig. 3, from which it is clearly shown that the intensity throughout the channel increases in the flow-on state, the more so at higher flow rates and closer to the inlet, up to 45% at the inlet for the highest flow rate of 6 mL/min (shear rate ~9 s$^{-1}$) that we consider here. Note, however, that the intensity also increases significantly, by about 10%, at the outlet at the lowest flow rate of 1 mL/min (shear rate ~1.5 s$^{-1}$). With these flow rates, we capture a laminar flow with corresponding Reynolds numbers between ~5 and ~25. For each flow rate, the increase of the intensity upon applying flow varies monotonically between inlet and outlet. To the best of our knowledge, other spectroscopic flow experiments conducted so far only varied the flow rate in ranges where no dependency has been observed[4,26].

In summary, the v-SFG data reveal that flow triggers an increase of surface charge at every position in a channel. This increase varies monotonically along the channel length so that it is largest at the inlet and smallest at the outlet. This change in surface charge upon flow is reversible. Additionally, we found that the surface charge variation is also a function of the flow rate. To explain our observations quantitatively, we need a model that combines microscopic surface chemistry and macroscopic processes such as advection and diffusion in a flow-system. This goes beyond the qualitative explanation by LIS et al.[4]. In the following section, we will introduce a self-consistent model that extends that of LIS et al.[4] and describes not only the flow-dependent surface charge but also the dependency on flow rate and position.

**Model of Surface Chemistry**

Here we will briefly discuss the qualitative features of our model. We incorporate two reactions that take place at the $CaF_2$-water interface (i.e., the wall of the flow channel): (i) The surface charging reaction of equation (1) in which only $F^-$ desorbs from the surface to leave behind a positively charged $CaF^+$ unit on the surface, together with its back-reaction that involves $F^-$ adsorption. This reaction is responsible for the surface charging, and has been considered responsible for the flow-induced change of the surface charge in previous work[4]. However, as demonstrated below, a second reaction is necessary to explain the observations: (ii) The dissolution reaction of equation (2) in which a neutral $CaF_2$ unit dissolves from the surface in the form of three dissolved ions (one $Ca^{2+}$ and two $F^-$), together with its back-reaction, i.e., precipitation. This reaction is the source of $F^-$ ions without changing the surface charge. The reactions are given by:

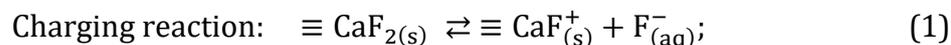
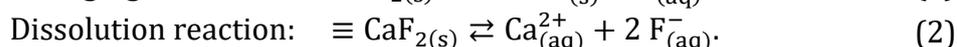

$$\text{Charging reaction:} \quad \equiv CaF_{2(s)} \rightleftarrows \equiv CaF^+_{(s)} + F^-_{(aq)}; \quad (1)$$
$$\text{Dissolution reaction:} \quad \equiv CaF_{2(s)} \rightleftarrows Ca^{2+}_{(aq)} + 2\,F^-_{(aq)}. \quad (2)$$

Both reactions have a fluoride ion in common, which couples them. Fig. 4a illustrates this view of the surface chemistry. While LIS et al.[4] and XI et al.[26] only considered the charging reaction from equation (1) to qualitatively explain the flow-induced changes in v-SFG spectra, we argue that a consistent and quantitative description of the experimental results also requires to incorporate the dissolution as described by equation (2). The charging reaction of equation (1) alone could not serve as a steady-state source for fluoride ions. During the continuous dilution with fresh solution from the reservoir in the flow-on period, the surface charge and thus the v-SFG signal would have to keep increasing, which is not observed experimentally as a steady-state v-SFG signal is reached after about 5 minutes. Therefore, our model considers the $CaF_2$ surface not only as an adsorption-desorption site for fluoride anions but also as an ion source for charge-neutral dissolution and precipitation of $Ca^{2+}$ and two $F^-$ according to equation (2). In fact, the dissolution reaction is not only a theoretical necessity but is also supported by the presence of $Ca^{2+}$ cations in flow experiments with $CaF_2$ [4].

The appearance of a surface charge gradient upon flow can be explained when considering a flow-induced gradient in fluoride concentration along the mineral surface, as illustrated in Fig. 4 b. This gradient forms when the laminar flow transports dissolved fluoride anions along the interface, which leads to a build-up of fluoride ions at the outlet and a positive gradient in fluoride concentration in the flow direction. Moreover, because the solution flowing into the channel from the reservoir has a lower fluoride concentration, the fluoride ions are also diluted by the fresh solution, the more so the closer to the inlet. Thus, flow reversibly lowers the fluoride concentration everywhere in the channel, yet more so closer to the inlet. As the surface charge increases with lower fluoride concentration according to equation (1), the surface charge becomes larger at positions closer to the inlet. This qualitatively explains the gradient of the increase in surface charge along the surface upon flow and thus in the v-SFG response.

**Numerical Modelling to Quantitatively Reproduce the Observations**

We model the channel as a cylinder of length $2L$ and radius $R$, and introduce the radial and axial coordinate $r$ and $x$, respectively. The charged and dissolving mineral surface of the channel is located at $r = R$ for $x \in [-H, H]$ (with $H < L$). The channel inlet and outlet are at $x = \pm L$, and $|x| \gg L$ denotes the interior of the two reservoirs in which we specify bulk concentrations and bulk pressure. The model geometry is depicted in Fig. 5a. Throughout, we will assume azimuthal symmetry and consider water to be an incompressible fluid with viscosity $\eta$ and dielectric constant $\varepsilon$ at fixed room temperature $T$. We are interested in finding the areal density $\sigma(x)$ of charged surface groups on the mineral surface, the fluid velocity $\mathbf{u}(x, r)$ in the channel, and the concentrations $\rho_i(x, r)$ of ion species $i = H, Cl, Na, Ca, F$ of which, however, only calcium and fluoride will turn out to be relevant. The system is solved for steady-state without any explicit time-dependence. The fluid flow is driven by an imposed static pressure drop between the two reservoirs such that the pressure profile $P(x, r)$ is also to be determined. We model the surface chemistry in terms of the dynamic Langmuir equation

$$\partial_t \sigma = k_{\text{des}}(\Gamma - \sigma) - k_{\text{ads}} \rho_F(x, R) \sigma \tag{3}$$

that accounts for the charging reaction of equation (1). Here $k_{\text{des}}$ and $k_{\text{ads}}$ denote the desorption and adsorption rate constants of fluoride, respectively, and $\Gamma$ denotes the surface density of chargeable groups. The surface charge density is thus denoted by $e\sigma(x)$, with $e$ the proton charge. In the steady-state conditions of interest here, $\partial_t \sigma = 0$, we can solve the dynamic Langmuir equation (3) to obtain the Langmuir isotherm

$$\sigma(x) = \frac{\Gamma}{1 + \frac{k_{\text{ads}}}{k_{\text{des}}} \rho_F(x, R)}. \tag{4}$$

Note that when the surface charge is in steady-state, $\partial_t \sigma = 0$, there is no net source of charge directed into the solution, and hence the flux of calcium ions $\mathbf{J}_{\text{Ca}}$ equals twice the flux of fluoride ions, $2\mathbf{J}_F = \mathbf{J}_{\text{Ca}}$. We should realize that the fluoride concentration at the surface needs to be determined and depends not only on the position $x$ but also (implicitly) on the flow velocity $\mathbf{u}$. The profiles of the ionic concentrations, the pressure, and the fluid velocity follow from the advection-diffusion equations (equation (5)) for ionic transport and the incompressible Stokes equation (equation (6)) for the fluid velocity $\mathbf{u}$,

$$\partial_t \rho_i = \boldsymbol{\nabla} \cdot (D \boldsymbol{\nabla} \rho_i - \rho_i \mathbf{u}) = 0; \tag{5}$$
$$\partial_t \mathbf{u} = \eta \nabla^2 \mathbf{u} - \boldsymbol{\nabla} P = 0; \quad \boldsymbol{\nabla} \cdot \mathbf{u} = 0. \tag{6}$$

Here $D$ is the diffusion constant that is assumed to be equal for all ionic species. It is computationally not feasible at the cm length scales of the experiment to resolve the electric double layers at the solid-liquid interface on the nm scale. Hence, we neglect a conduction term to equation (5) and a coupling to the Poisson equation that would be required to account for electrostatics. Given that the concentration profiles of sodium, hydrogen, and chloride ions have no sources and sinks and are not affected by conduction within this approximation, we can assume them to be spatially constant, with values set by the reservoir. Moreover, since $\partial_t \sigma = 0$ and $\mathbf{J}_{Ca} - 2\mathbf{J}_F = 0$, we can reduce the number of concentration profiles further by imposing local charge neutrality and hence setting $\rho_F = 2\rho_{Ca}$ in the simplest version of the theory that we focus on here, in the main text. However, in Supplementary Discussion 1 we go beyond the assumption of local charge neutrality and allow for electric double layers by solving a similar, but orders of magnitude smaller, system with a conduction term and the Poisson equation; we find that the non-electrostatic version of the theory captures the key physics. To solve for the fluoride concentration, we seek solutions to equations (5)-(6) that satisfy the following boundary conditions. Deep in the two reservoirs, we impose identical ionic bulk concentrations $\rho_{i,b}$ for each species in the reservoir, and a pressure that differs by an imposed pressure drop $\Delta P$. On the solid channel wall we apply no-slip boundary conditions for the flow velocity and vanishing normal fluxes for fluoride for $H < |x| < L$. On the dissolving wall, at $|x| < H$, we impose the dissolution and precipitation reaction (2), resulting in $\mathbf{n} \cdot \nabla \rho_F(x, R) = \mathbf{J}_F(x)/D$, where $\mathbf{J}_F(x)$ is the normal flux of fluoride ions into the solution given by

$$\mathbf{J}_F(x) = 2\, k_{\text{dis}} \left(1 - \frac{k_{\text{prec}}}{2\, k_{\text{dis}}} \rho_F^3(x, R)\right). \tag{7}$$

Here we introduced the surface normal $\mathbf{n}$ pointing into the solution, the dissolution rate-constant $k_{\text{dis}}$ of the $CaF_2$ units, and the precipitation rate-constant $k_{\text{prec}}$, in line with the reaction of equation (2); the prefactor 2 is a stochiometric constant resulting from the fact that two F⁻ ions go into the solution when one $CaF_2$ unit dissolves. The ratio of the rate constants defines the saturation concentration $\rho_{F,\text{sat}} = (2\, k_{\text{dis}}/k_{\text{prec}})^{1/3}$ upon which no further dissolution occurs. Note that the local concentrations govern the net precipitation rate at the surface, which couples to the flow through the advection term in equation (5), such that a flow-dependent surface charge is obtained if the dissolution and charging reactions share an ion. For $CaF_2$ the shared ion is F⁻.

Due to dissolution, the net flux of ions into the flow channel creates, even without flow, a reaction-diffusion equilibrium concentration profile that extends over the length of the channel and into the two reservoirs, as illustrated in Fig. 5a. With flow, a lateral and a normal gradient of fluoride ions develop because the fluid flow increases the rate of transport out the channel, the more so on the central axis of the channel where the flow velocity is largest, as is evident from Fig. 5b. The radial concentration profile near the surface is well described by diffuse boundary layer theory (Supplementary Discussion 3), and has a thickness $\delta(0) \approx 0.4$ mm at the center ($x = 0$) of the channel, and gets thinner/thicker upon approach of the inlet/outlet following approximately the scaling $\delta(x) \propto (x + H)^{1/3}$. Because the diffuse boundary layer is localized at the surface where the fluid velocity is much smaller than the channel-averaged velocity, the effective transport rate here is accordingly lower, allowing for a significant concentration to remain close to the surface.

## Concentration, Surface Charge, and Dissolution Profiles

We have numerically calculated the fluoride concentration profiles and fit the experimental v-SFG results using experimental and literature values for the model parameters (Fig. 6). We used $k_{dis} = 0.027\ \mu\text{mol m}^{-2}\text{s}^{-1}$, [30,31], $\rho_{F,sat} = 10\ \rho_{F,b} = 10\ \mu M$, [32], $k_{ads}/k_{des} = 10\ \mu M^{-1}$, and $\Gamma = 10\ \text{nm}^{-2}$. The equilibrium constant $k_{ads}/k_{des}$ and the density of chargeable groups $\Gamma$ are not directly reported in the literature. However, the combination chosen here is not only in agreement with the experimental equilibrium surface potential of about 70 mV (if the electric double layer is taken into account)[32-34], but is also in agreement with the observed magnitude of typical flow-induced changes of the zeta potential ($\approx 15$ mV) as reported in streaming potential measurements[15]. The dissolution rate constant $k_{dis}$ is low and roughly corresponds to one monolayer of the surface dissolving every hour, as expected for a poorly soluble mineral[30,31]. Fig. 6a shows the comparison between the experimentally observed changes in v-SFG intensity as a function of flow rate and position and those found in our numerical calculations (in which we tune the pressure drop to fit the experimental flow rate). The overall increase in the v-SFG response with increasing flow rate and its decrease from the inlet to the outlet is very similar to the characteristics of the surface charge (Fig. 6b). The surface charge increases by 50-100% compared to no-flow conditions, and varies by 10-20% laterally. The corresponding change in the surface potential $\psi$ (as calculated by the Gouy-Chapman relation $\psi = 2\frac{k_B T}{e} \sinh^{-1}(2\pi\sigma\lambda_B\lambda_D)$) is significantly lower in this nonlinear screening regime where $\psi$ exceeds the thermal voltage of 25mV, differing by only around 20% between the flow-on and flow-off states. Here $\lambda_B = 0.7$ nm is the Bjerrum length of water at room temperature and $\lambda_D = 7$ nm the Debye length for a solution of 1 mM of NaCl and pH 3. Finally, to estimate the change in the SFG intensity, we use the approximation $(I_{on} - I_{off})/I_{off} = (\psi_{on}/\psi_{off})^2 - 1$, as discussed in Supplementary Note 1.

The validity of the Gouy-Chapman relation is verified in the Supplementary Discussion 1, where we show that the curvature of the channel wall on electrostatics can be neglected, which is expected as the Debye length is much shorter than the channel radius, $\lambda_D/R \ll 1$. This also justifies approximating the dissolving surface, which is a flat plate in the experiments, by a cylindrical channel wall in our numerical calculations. Similarly, the diffuse boundary layer and the electric double layer do not affect each other as their thicknesses differ by orders of magnitude, $\lambda_D/\delta \approx 10^{-4}$.

The calculated interfacial fluoride concentration (Fig. 6c) decreases with increasing flow rate and increases from inlet to outlet. The reverse relation between fluoride concentration and surface charge stems from the Langmuir isotherm (Equation (4)). Interestingly, although the fluoride concentration increases in the flow direction, Fig. 6c shows that the flow-induced dilution ensures that the concentration is lower throughout the channel than in the no-flow state. The concentration at the boundary of the dissolving area, at $x = -H$, is equal to the bulk concentration $\rho_{F,b}$ in flow-on conditions, beyond which it rises rapidly and then slowly reaches a maximum at $x = H$. This rapid increase of the fluoride concentration at $x = -H$ is due to the locally low precipitation rate compared to upstream locations near $x = H$.

In the no-flow state, the fluoride concentration in the center corresponds to the saturation concentration $\rho_{F,sat}$. Without a precipitation reaction, the concentration would even be orders of magnitude larger. In our calculations that describe the experiment, the flow-induced change in fluoride concentration is on the order of ~30% of the saturation concentration ($\rho_{F,sat} = 10$ µM), as is evident from Fig. 6c. In an absolute sense, this change of concentration is orders of magnitude lower than the background ionic strength (~2 mM). While the change in F- concentration is sufficient to affect the surface charge, it is negligible for the total ionic strength that governs the screening of the surface charge. Thus, we can conclude from our combined experiments and theory that the flow-induced changes in the v-SFG signal are connected to the surface charge itself rather than changes in charge screening.

Finally, we draw attention to Fig. 6d, which reveals that the net dissolution rate also exhibits a gradient along the mineral surface and increases with flow rate. Similar to the position- and flow-rate-dependent shift in surface charge, the position-dependent net dissolution rate is also a result of the heterogeneous concentration profile. More importantly, however, we can see that the dissolution rate increases by orders of magnitude from flow-off to flow-on conditions. In the flow-off state, the surface concentration is nearly equal to the saturation concentration $\rho_F(x, R) \approx \rho_{F,sat}$, which causes the net dissolution rate in flow-off conditions to be ~$10^5$ times smaller at the center than the maximum dissolution rate $2k_{dis}$. During flow the dissolution is almost as large as $k_{dis}$. The net dissolution rate thus increases by ~5 orders of magnitude due to flows of ~mL·min-1 (or shear rates 1.5-9 s$^{-1}$), from effectively $10^{-7}$ to $10^{-2}$ µ·m-2·s-1.

To investigate the universal scaling behavior of the fluoride concentration with channel geometry, dissolution, and flow rate, we derive in Supplementary Discussion 2 an analytic expression for the radially averaged fluoride concentration profile $\bar{\rho}_F(x)$ under the assumption that the concentration profile is independent of the radial coordinate (such that $\rho_F(x, R) = \bar{\rho}_F(x)$) and upon neglecting precipitation. Key parameters of this expression are the Peclet number $\text{Pe} = 2\bar{u}L/D$, where $\bar{u}$ is the channel-averaged fluid velocity, and the dimensionless concentration difference $\Delta\rho_F = k_{dis}L^2/DR\rho_{F,b}$ between the channel center and the reservoir under flow-off conditions. Here Pe captures the importance of advective transport relative to diffusive transport, and $\Delta\rho_F$ is a measure for the dissolution rate. In Fig. 7a, we plot our analytic expression of $\bar{\rho}_F(x)/\rho_{F,b}$ for the geologically relevant case $\rho_{F,b} = 10^{-6}$ mol·L-1 and $\Delta\rho_F = 3.6$ for several values of Pe, yielding remarkable agreement with our numerically exact solution of equations (5)-(7) for the long-channel limit. Interestingly, the parabola-like low-Pe regime and the linear-like high-Pe regime of the graphs in Fig.7a are borne out by the limiting cases of our analytic expression, which are given by

$$\frac{\bar{\rho}_F(x)}{\rho_{F,b}} = \begin{cases} 1 + \Delta\rho_F \left(1 - \frac{x^2}{L^2}\right) & \text{if } \text{Pe} \ll 1; \\ 1 + \frac{\Delta\rho_F}{\text{Pe}}\left(1 + \frac{x}{L}\right) & \text{if } \text{Pe} \gg 1. \end{cases} \quad (8)$$

This expression makes explicit that all excess fluoride gets washed out of the channel for $\text{Pe} \gg 1$, the more so at the inlet ($x = -L$) than at the outlet ($x = L$). In Fig.7b, we plot the fluoride concentration at the center of the channel ($x = 0$) in the full range of Pe, which reveals a diffusion-dominated ($\text{Pe} \ll 1$) and an advection-dominated ($\text{Pe} \gg 1$) regime with Pe-independent limiting fluoride concentrations; a significant dependence on Pe is only found in a rather narrow crossover regime at $\text{Pe} \approx 1$. This narrow Pe regime could explain why previous authors did not observe a change in surface charge when the flow rate was varied [4,7,9,26], as there is little flow dependency in the high Peclet regime ($\text{Pe} \gg 1$). Note that if the radial concentration profile is not constant, but rather described by a diffuse boundary layer with thickness $\delta \ll R$ (as is the case for our experiments), the transition between flow and no-flow states is expected to broaden significantly and to occur at higher flow rates as the Peclet number has to be replaced by the Sherwood number (Sh $\propto$ Pe$^{2/3}$), which we discuss in Supplementary Discussion 3.

How an almost insoluble mineral such as CaF$_2$ can give rise to a significant concentration difference ($\Delta\rho_F > 1$) in our experiment can be understood by interpreting $\Delta\rho_F$ as a ratio of the time scale $\tau_{\text{dif}} = L^2/D \approx 10^5$s for a fluoride ion to diffuse from center to outlet and the dissolution timescale $\tau_{dis} = \rho_{F,b}R/k_{\text{dis}} \approx 10$ s by which dissolution can significantly change the bulk concentration. While the typical time for the concentration to change by dissolution is slow, the timescale by which diffusion can smooth out this concentration change is even slower in the present case of a long macroscopic channel. For this reason, the slow dissolution is offset by the even longer diffusion time, leading to a significant concentration difference ($\Delta\rho_F = \tau_{\text{dif}}/\tau_{\text{dis}} > 1$). Note that the dissolution time is determined by the surface-to-volume ratio of the channel and is hence linear in the radius $R$, while the diffusion time is quadratic in the channel length $L$. Thus the dimensionless concentration difference $\Delta\rho_F$, which needs to be of order unity or larger for a flow-dependent surface charge and dissolution rate, is strongly dependent on both the absolute length $L$ as well as the aspect ratio $L/R$ of the channel. In fact, in the limit of large channels, $\Delta\rho_F$ inevitably becomes so large that the assumption of a negligible precipitation reaction is invalidated. Moreover, the inverse scaling of $\Delta\rho_F$ with the bulk fluoride concentration $\rho_{F,b}$ explains the experimental finding of Ref. (4) that there is no flow-dependent surface charge when an excess of fluoride is added. We experimentally reproduced this behavior for a 4 mM fluoride bulk concentration (Supplementary Figure 3a) and confirm that under these conditions no surface charge gradient was observed (Supplementary Figure 3b).

## Discussion

Our theoretical model and v-SFG experiments can be transferred to other interfaces to determine the charging reaction mechanism and dissolution reaction mechanism. In the present case of calcium fluoride, the surface charging reaction (equation (1)) is desorptive, which implies that a decrease in the fluoride concentration increases the surface charge. Meanwhile, the dissolution reaction (equation (2)) is a source of reactive ions and hence increases the fluoride concentration. As flow brings the concentration closer to the reservoir's bulk concentration, the surface charge increases, more so at the inlet than at the outlet. The sign of the charges (i.e., whether an anion or cation is released) does not play a role in the surface charge's flow dependency. Interestingly, the combination of the surface being charged by desorption and dissolution being a source of ions partaking in this charging reaction is only one of several ways to establish a flow-dependent surface charge. For instance, a precipitating surface in contact with an oversaturated solution is a sink for the reactive ions. In this case, the concentration would increase upon flow, which for a desorptive charging mechanism would lead to a decrease, rather than an increase, in surface charge closer to the inlet. Another charging mechanism is the adsorption of a reactive ion, where the surface charge increases with increasing concentration of reactive ions. This causes the surface charge to increase upon flow when the surface is a sink and to decrease upon flow when the surface is a source of reactive ions. Note that whether the surface is a sink or a source of ions is not necessarily determined by whether the mineral is dissolving or precipitating. For instance, protons play a role in the charging of iron oxide surfaces and are consumed when iron oxide dissolves[35]. In this case, the surface is a sink below the saturation concentration and a source above saturation concentration. As protons are often consumed in dissolution and partake in the charging of metal oxides, we expect such counter-intuitive behavior to be relatively common.

We summarize the expected change in surface charge upon flow for different combinations of charging and dissolution reactions in Fig. 8. If either the dissolution or charging reaction is known, the other reaction mechanism can be established by inspecting the sign of the v-SFG change upon flow or the slope of the surface charge gradient. This could be useful for the investigation of other minerals, for instance, the notoriously complex surface chemistry of silica[11,36-40].

Next we discuss in what natural setting flow-dependent charge and dissolution could be relevant, considering a common geological system. While we have focused on the position-dependent surface charge, as this is a quantity indirectly measured by v-SFG, the dissolution flux of equation (7) also depends on the fluoride concentration. Therefore, our model directly and naturally implies that also the dissolution flux is flow- and position-dependent. This insight brings us to discuss the more general case of mineral dissolution in geological environments, particularly whether and where such a dissolution rate gradient can be expected. As already emphasized, the gradients of interfacial concentrations can only be significant if there is a sufficiently large relative concentration difference ($\Delta \rho_F \gg 1$) between the reservoir and the channel (or porous material) without flow. At first sight, this requires high dissolution rates. However, also the geometry and the channel dimensions play a role, and we have seen that macroscopic and elongated channels are favorable for a large $\Delta \rho_F$. These conditions are met naturally in porous networks such as rocks and soils. In addition to the condition of a significant concentration difference, also a significant fluid velocity, or Peclet number, is required.

Using our analytic results (equation 8), we estimate that in pores of centimeter lengths ($L = 10^{-2}$ m) and sub-millimeter widths ($R = 10^{-4}$ m) and for fluid velocities of $\bar{u} =10^{-5}$ m·s$^{-1}$ and dissolution rate constants that exceed $k_{\text{dis}} = 10^{-11}$ mol·m$^{-2}$ s$^{-1}$) , there will be a significant surface charge gradient. We note that these dissolution rates are typical for silicate minerals[28,29], and these flow rates are typical for water flow through soils[41,42]. We compare numerical and analytic results for these parameter values at different flow rates in Fig. 7b. The concentration profile changes dramatically from a no-flow state to a flow velocity of $\bar{u} = 10^{-5}$ m·s$^{-1}$ corresponding to Pe $\approx 10^2$. In the case that the concentration in the pore becomes close to the saturation concentration ($\Delta\rho \approx \rho_{\text{sat}}/\rho_{\text{F,b}}$), there will also be a flow-rate and position-dependent dissolution rate. Such a flow-dependent dissolution rate has been suggested to explain commonly observed dependencies of dissolution rate with pore size[29]. Therefore, our experimental findings and numerical models can be expected to be of geological relevance. Additionally, our approach of combining the model and position-resolved v-SFG experiments itself could be used to investigate charging and dissolution mechanisms, which are important for soil remediation[9] and industries such as froth flotation [6].

In conclusion, we demonstrated that the surface charge and dissolution rate of calcium fluoride show macroscopic gradients along the interface when flow is applied. This was shown by the combined use of surface-specific v-SFG spectroscopy and full numerical calculations of reaction-diffusion-advection equations, including the interfacial chemistry at water-mineral interfaces. We extended v-SFG spectroscopy spatially to investigate the surface chemistry of minerals upon flow on a macroscopic scale. The observed gradient in the v-SFG response can be correlated to a gradient in the surface charge upon flow. A dissolution-diffusion-advection process can entirely explain such a gradient. The key physicochemical mechanism is captured by the coupling of a dissolution reaction with flow and surface charge. This dissolution reaction creates a steady-state concentration profile of fluoride ions, which depends on both position and flow rate. Both the surface charge and dissolution rate are influenced by the local fluoride concentration. Thus, dissolution also causes a position-dependent dissolution rate and surface charge during flow. Our reaction-diffusion-advection model can be generalized to arbitrary surface and dissolution reactions, and we expect that the method of position resolved v-SFG spectroscopy can be transferred to other systems in order to determine charging and dissolution mechanisms. Interfaces exhibiting both dissolution and charging in flowing water occur naturally in a wide variety of systems, and only minor confinement is needed to ensure that both surface charge and dissolution rate depend on position and flow rate. Thus, our finding of a flow-induced gradient in the dissolution reaction is expected to impact geological research of porous materials. Also, scientific disciplines that rely on the use of surface potentials, such as electrokinetics, microfluidics, and nanofluidics, or those involving measurements of surface potentials may be interested in its dependency on the position and flow rate.

## Methods

### SFG spectroscopy

The SFG spectra presented in this work are recorded with an experimental setup based on a Ti:sapphire amplifier (Solstice® Ace™, Spectra Physics) that generates 800 nm pulses with 1 kHz repetition rate and ~40 fs duration. For producing broadband infrared pulses (full width at half maximum, FWHM ~400 cm$^{-1}$) with 4 µJ pulse energy, a commercial optical parametric amplifier (TOPAS Prime, Spectra Physics) is used together with a non-collinear difference frequency generation (NDFG) scheme. By guiding the visible upconverting pulse through a Fabry–Perot etalon (SLS Optics Ltd), it is spectrally narrowed to an FWHM of ~20 cm$^{-1}$ with ~15 µJ pulse energy. The spectral resolution of the SFG signal is obtained by using a spectrograph (Acton Spectro Pro® SP-2300, Princeton Instruments). An electron multiplied charge-coupled device (emCCD) camera (ProEM 1600, Roper Scientific) is used for detection. All spectra are recorded in ssp polarization combination (s-polarized SFG, s-polarized visible and p-polarized IR) and at incident angles $\theta_{vis} \approx 73°$ and $\theta_{IR} \approx 76°$ of the visible and infrared pulse, respectively.

### Chemicals

Sodium chloride was purchased from Carl Roth (>99.5%, CAS 7647-14-5), sodium fluoride from Merck (99.99%, CAS 7681-49-4), and concentrated hydrochloric acid from VWR (37 w% CAS 7647-01-0). All reagents were used as received without further purification.

### Sample preparation

The rectangular fluorite prism (dimensions 4 cm × 1 cm × 0.5 cm) purchased from EKSMA optics was baked at 500 °C for at least 2 h to remove organic residues. Before and after baking, it was rinsed with demineralized water, filtered with a Millipore unit (resistivity = 18MΩ cm). The salt was dissolved in a pH 3 solution of hydrochloric acid prepared from the concentrated acid. If not stated otherwise, a 1 mM NaCl solution at pH 3 was used.

### Flow Set-up

A scheme of the flow set up and a technical drawings of the flow cell are shown in Supplementary Figure 4. The design of the flow cell was optimized to ensure laminar flow. The prism is mounted on top of the flow channel. A 0.03 mm Teflon film is used for sealing. The flow rate was between 1 and 6 mL/min, corresponding to Reynold number of the order of ~5 − 25 and shear rate of ~1.5 to 9 s$^{-1}$, which ensures laminar flow over the investigated flow cell. A reservoir with ~100 mL of the solution is connected with the flow cell via Tygon® tubes (ISMCSC0832 MHSL 2001; inner diameter 4.8 mm, outer diameter: 8.0 mm). The peristaltic pump (Masterflex 77924-70 L/S Ethernet/IP Network-Compatible Pump) is controlled by a self-written LabView program. The same program controls the motors (Newport Spectra-Physics GmbH CONEX-TRA25CC) of the x- and y-stages (Manual Linear Stage, Crossed-Roller Bearings, 25.4 mm Travel, M6).

### Experimental Procedure

One stage is used to bring the center of the channel perpendicular to the flow direction into the focus of the incident laser beams. To do so, a prism with 100 nm gold coating in the area of the flow channel is mounted on the flow cell, and the stage is moved to the edges where the gold signal disappears. The center is chosen in the middle of the edges. The center of the flow cell in flow direction was marked with an engraving on the flow cell's surface. The second stage orthogonal to the first one is used to adjust the laser focus to the center of the flow cell in flow direction by eye with the help of the engraving. Afterward, the flow cell and the tubes are rinsed with MilliQ water. Before spectra are collected, the flow cell with the freshly cleaned prism mounted is flushed 2 minutes at a high flow rate of ~600 mL/min with the solution, followed by 10 minutes of rest to ensure equilibration of the interface. The second stage is used to change the position along the flow direction. At each position, spectra were recorded with an acquisition time of 5 seconds.

**Data Processing**

The spectra were subtracted by a background obtained by measuring a spectrum with a blocked IR beam for the same time as for the usual spectrum. Cosmic rays were identified by a self-written MatLab script once a tolerance for the difference between the intensities of neighboring pixels was exceeded. The cosmic ray was replaced by the average of the two previous points. Usually, a spectrum (1600 data points) contains less than 1 % of cosmic rays. Integration was done by summation of the background-subtracted intensities in the range of the IR-pulse. By inserting a polystyrene foil in the IR beam path prior to a gold sample, a spectrum with dips due to the infrared absorption of foil is obtained. The absorption dips were used for calibrating the wavenumbers. To calculate relative increases of the signal upon flow, spectra over ~ 5 min prior to starting the flow and spectra over ~ 5 min prior to turning the flow off are averaged. Several flow on/off cycles as well as single flow on/off cycles were taken into account for the presented data (see Supplementary Figure 1). From Supplementary Figure 1, it is also clear that the spectra chosen for averaging represent a steady-state.

**Numerical Calculations**

We model a system involving fluid flow, diffusion and two chemical reactions. The reactions are a desorption reaction responsible for the surface charging, and a dissolution reaction, responsible for the flux of reactive ions in the channel. Mathematically, the reaction-diffusion-advection problem is described by equations (3)-(7). We use the approximation of local charge neutrality ($\rho_F(x,r) = 2\rho_{Ca}(x,r)$ and $\rho_{Na}(x,r) + \rho_H(x,r) = \rho_{Cl}(x,r) = 2$ mM (a spatial constant)) to reduce the computational complexity. Note that NaF rather than CaF$_2$ was used in the experiments to fix the bulk fluoride concentration, which only introduces a minor discrepancy in the Ca$^{2+}$ concentration. The dissolution of CaF$_2$ is taken into account by the boundary condition for the dissolution equation,

$$\boldsymbol{n} \cdot \boldsymbol{J}_F(x) = 2\, k_{dis}\left(1 - \frac{k_{prec}\rho_F^3(x,R)}{2\, k_{dis}}\right) = D\partial_r\rho_F(x,r)\Big|_{r=R}. \tag{9}$$

We ignored electrostatics, the Poisson equation, and a conduction term in equation (4). All this would require a nm-scale resolution that is computationally intractable on the cm-scale of the channel. Small scale calculations, however, in which electrostatics is fully incorporated through Poisson-Nernst-Planck-Stokes equations, can be found in the Supplementary Discussion 1; they reveal essentially the same key physics as the charge-free case. Nevertheless, we can convert the areal density $\sigma(x)$ of CaF$^+$ groups on the surface to a surface charge $e\sigma(x)$ and to a surface potential $\psi(x)$ if we make use of the Gouy-Chapman relation (which is also validated in the Supplementary Discussion 1),

$$\psi(x) = 2\frac{k_B T}{e} \sinh^{-1}(2\pi\sigma(x)\lambda_B\lambda_D). \tag{10}$$

Here $e$ is the proton charge, $\lambda_B = e^2(4\pi\epsilon_0\epsilon_r k_B T)^{-1}$ is the Bjerrum length, $\lambda_D = (8\pi\lambda_B\rho_s)^{-1/2}$ the Debye length where the total ionic strength $\rho_s = 2$ mmol/L is determined by the background NaCl and HCl salt concentration in the experiments, $e$ is the electron charge, $\epsilon_0$ the electric permittivity of vacuum, $\epsilon_r$ the relative permittivity of water, $k_B$ the Boltzmann constant and $T$ the temperature. The Debye length $\lambda_D$ is spatially constant as the concentration of reactive ions is too low to change the total salt concentration.

The differential equations are solved numerically in the commercial finite-element software package COMSOL MULTIPHYSICS. The modeling geometry is described using cylindric coordinates $(x, r, \theta)$, as can be seen in Fig.6 and is assumed to be azimuthally symmetric. The origin is located at the center of the channel. The channel length runs from $x = -L$ to $x = L$, with $L = 20$ mm, and the reactive surface runs from $x = -H$ to $x = H$, with $H = 12.4$ mm. The channel is connected to two bulk reservoirs of length $L_b = 42$ mm. The radius of the channel is $R = 2.4$ mm. The diffusion constant $D = 10^{-9}$ m$^2$·s$^{-1}$ and viscosity $\eta = 10^{-3}$ Pa s were used. A sketch of the model geometry can be seen in Fig. 5a. As boundary conditions, we impose bulk concentrations $\rho_i(-L_b) = \rho_i(L_b) = \rho_{b,i}$ at the reservoir in- and outlet with $\rho_{b,F} = 1$ µM, a pressure difference $P(L_b) - P(-L_b) = \Delta P$ over the channel, no-slip boundary conditions at the channel wall, $\mathbf{u}(R) = 0$, and no-flux boundary condition at the edge of the channel without a reactive surface $\mathbf{n} \cdot \mathbf{J} = 0$, where $\mathbf{n}$ is the surface normal. Dissolution is responsible for a net flux of ions out of the channel wall in to the channel in the region $-H < x < H$ as given by equation (7).

The steady-state solution for the surface charge is given by the Langmuir isotherm $\sigma(x) = \Gamma(1 + k_{\text{ads}}\rho_F(R, x)/k_{\text{des}})^{-1}$ and the solution for the lateral component of the velocity profile is given by Poiseuille flow $u_x(r)/u_{\max} = 1 - (r/R)^2$, with $u_{\max} = R^2\Delta P/8\eta L$.

## Data Availability

The data that support the findings of this study are available from the corresponding authors upon reasonable request.

## Acknowledgement

P.O. gratefully acknowledges financial support by the Max Planck Graduate Center with the Johannes Gutenberg University Mainz (MPGC). All authors acknowledge the technical support of Marc-Jan van Zadel particularly the design of the flow cell. All authors thank anonymous Reviewer 2 for pointing us to relevant literature on diffuse boundary layer theory. This work is part of the D-ITP consortium, a program of the Netherlands Organisation for Scientific Research (NWO) that is funded by the Dutch Ministry of Education, Culture and Science (OCW).

## Author Contribution

P.O., E.H.G.B. and M.B. designed the experimental part of the research project and provided a qualitative interpretation of the experimental results. P.O. performed the experiments and analyzed the data. W.Q.B., M.D and R.v.R. provided a quantitative interpretation of the experimental results. W.Q.B. performed the numerical and analytic calculations and generalized the findings. All authors discussed the results and wrote the manuscript.

## Competing interests

The authors declare no competing interests.

# Figures

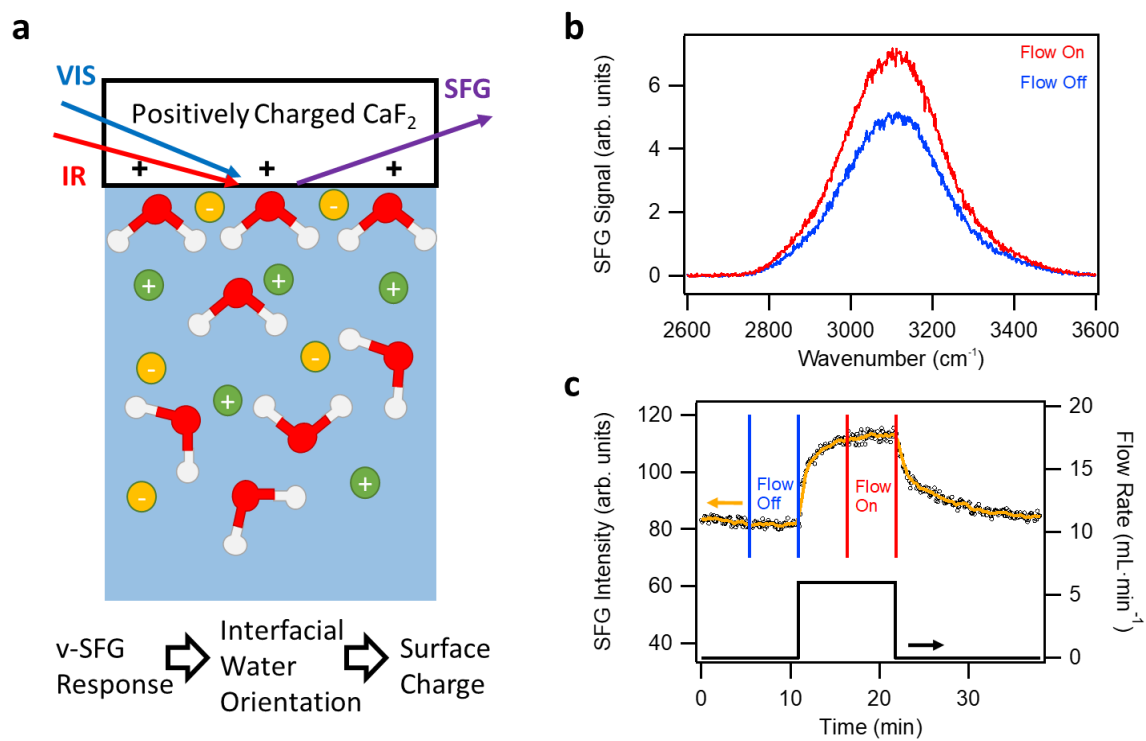

**Fig. 1. Studying the CaF$_2$-water interface by v-SFG.** (a) Illustration of the measurement method. In vibrational sum frequency generation (v-SFG) spectroscopy, a visible (Vis) and infrared (IR) pulse overlap in space and time at the interface of interest, generating radiation at the sum of the frequencies (SFG). If an IR pulse is in resonance with the OH stretch vibration of water molecules, the obtained v-SFG spectra provide information on the orientation and polarization of interfacial water molecules. The illustration shows water molecules for the first few hydration layers as well as a schematic distribution of arbitrary ions (anions in yellow, cations in green). At the charged CaF$_2$-water interface, the orientation and polarization of the interfacial water molecules is determined by the surface charge. Thus, the v-SFG signal can be used as an indirect measure for the surface charge. (b) v-SFG spectra in arbitrary units (arb. units) in the OH stretch region of the CaF$_2$- water interface at pH 3 (1mM HCl and 1 mM NaCl) under flow-off conditions (blue) and flow-on conditions (red). (c) Time trace of integrated SFG spectra (black circles) at the CaF$_2$-aqueous solution interface at pH 3 (1mM HCl and 1 mM NaCl) with one flow on-off cycle (black curve). Each circle represents one spectrum, integrated between 2650 and 3600 cm$^{-1}$. The solid orange line is a ten-point moving average to guide the eye. The vertical blue and red lines highlight the steady-state regimes for flow-off and flow-on conditions. As the intensities are hardly changing over time, a steady-state can be assumed.

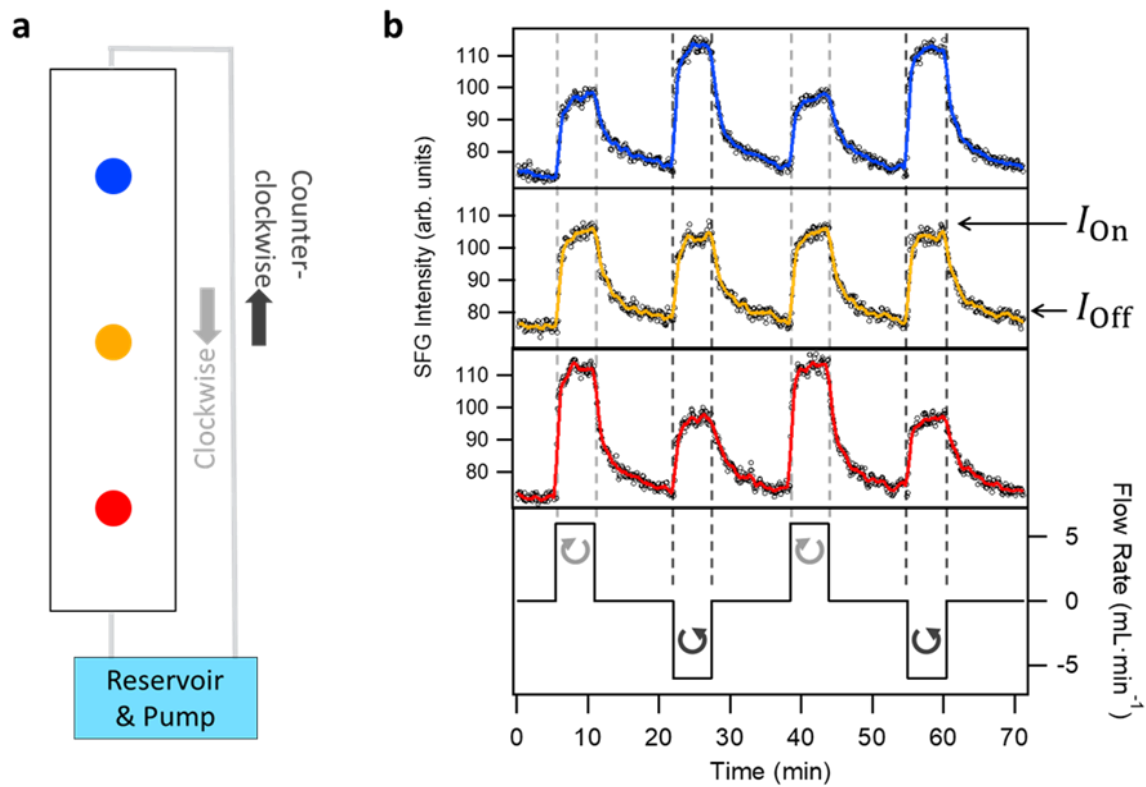

**Fig. 2**. **Position and flow direction-dependent change in the v-SFG response upon flow.** (a) Schematic representation of the flow set up. The aqueous solution reservoir (~100 mL of a 1 mM NaCl and HCl (pH 3) solution) is pumped via tubes and the flow cell using a peristaltic pump. The flow channel is 24.8 mm long, 4.3 mm deep, and 4.8 mm wide. The flow direction can be changed from clockwise to counterclockwise. The v-SFG spectra were recorded at three spots in the channel. One spot in the center (yellow) and two points separated by 8 mm from the center in the two directions of the flow channel (blue and red, respectively). (b) Time trace of integrated SFG spectra (black circles) at the interface of the solution and calcium fluoride for several flow on-off cycles (black curve). As indicated by the arrows and the different signs of the flow rate, different flow directions were employed. The colored solid lines are ten-point averages to guide the eye and highlighting at which spot in the flow channel the spectra were recorded. For the time trace measured at the center, the intensity levels of the spectra with ($I_{on}$) and without flow ($I_{off}$) are highlighted by arrows.

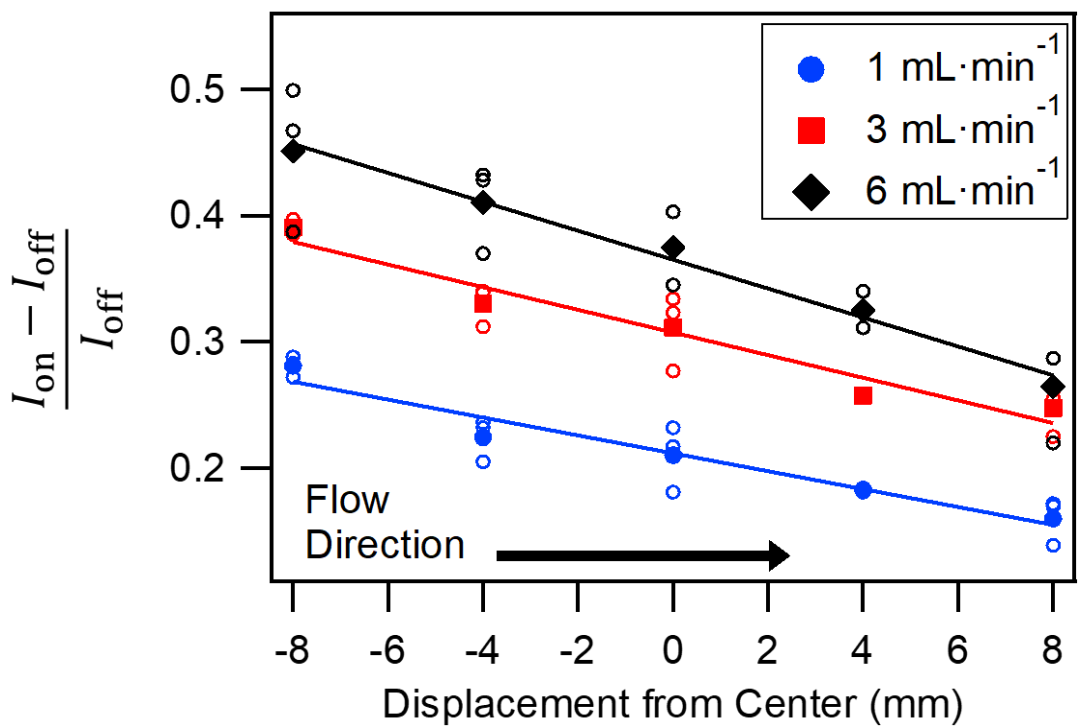

**Fig. 3. Gradient in the increase of the v-SFG intensity for different flow rates.** At different positions, the relative increase in the v-SFG intensity at the CaF$_2$-water interface is calculated based on averaged steady-state spectra under flow-on and flow-off conditions as marked in Fig. 1 c and Supplementary Figure 1, for the three indicated flow rates. The open circles represent the relative increase upon flow and the filled symbols the mean at their corresponding positions. The error of the individual ratios (open circles) due to small fluctuations within the steady-states of one flow-on/off cycle (see Fig1c of the main text and Supplementary Figure 1) is negligible compared to the spread. The solid lines are guides to the eye. The solution used was a 1 mM NaCl solution in pH 3 HCl with an additional 1 μM NaF. The addition of a very low NaF concentration constrains the number of unknown parameters in our model. Comparing the black data points here with the data points in Supplementary Figure 2 see shows that the added 1 μM NaF does not noticeably influence the interfacial events. The position is given as displacement from the center, with values increasing in the direction of flow, indicated by the arrow.

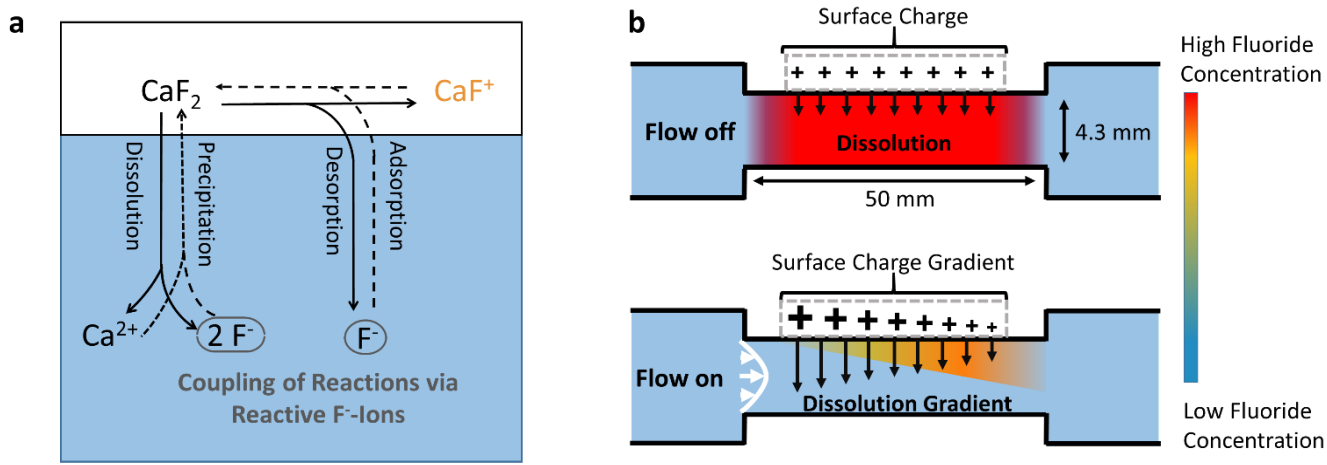

**Fig. 4**. **Model of Surface Chemistry along the flow channel.** (a) Illustration of the two interfacial reactions. $CaF_2$ is not only an adsorption-desorption site for fluoride anions (equation (1)), which alters the surface charge, but also experiences charge-neutral dissolution-precipitation processes (equation (2)). The shared fluoride anion couples the two reactions. (b) Illustration of the flow channel. Without flow (upper part), dissolution creates an enhanced F⁻ concentration in steady-state, constant over the dissolving mineral surface, if the concentration reaches the saturation concentration. Dissolution rates and surface charge are also homogeneous along the mineral's surface. However, upon flow (lower panel), the concentration is lowered everywhere in the channel, the more so closer to the inlet. This leads to an increase in the surface charge and higher net dissolution than in no-flow conditions, both becoming smaller at the outlet.

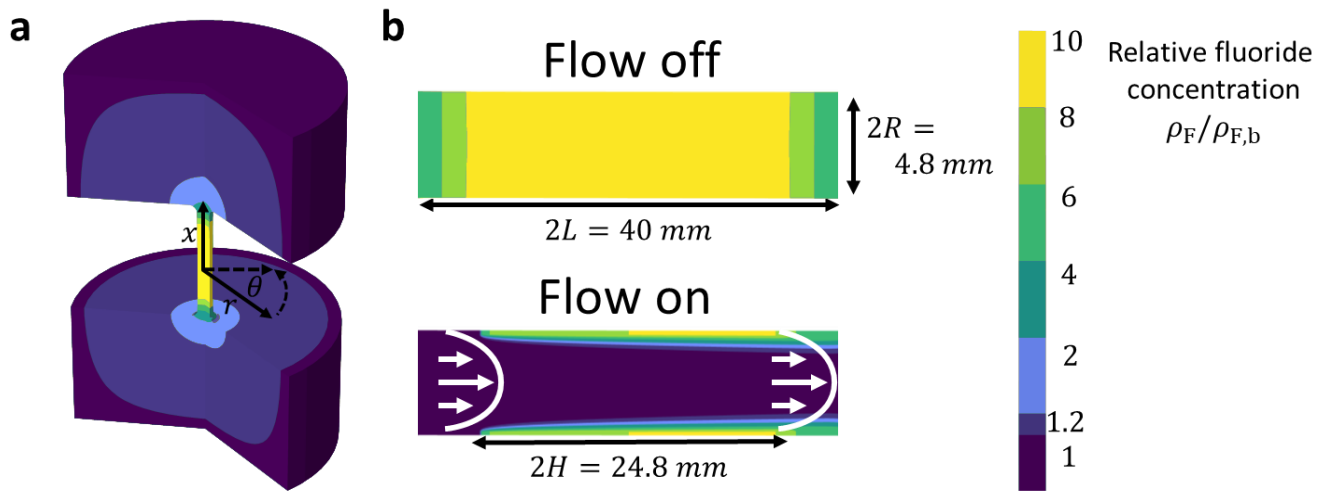

**Fig. 5. Channel geometry and numerical solutions to the reaction-diffusion-advection equations (3)-(7).** Illustration of the flow channel as used for the numerical calculations in COMSOL with resulting concentration profiles. The center of the flow channel is at the origin ($x = 0, r = 0$). The channel length and diameter are $2L = 40$ mm and $2R = 4.8$ mm, respectively, and the dissolving surface length is $2H = 25$ mm. The color represents the fluoride concentration. In (a) the result of the numerical calculations in the full three-dimensional model geometry can be seen in flow-off conditions. Panel (b) is a zoom in on the channel under flow-off conditions (top) and flow-on conditions (bottom) at 1 mL·min$^{-1}$ in the positive $x$-direction as also indicated by the parabolic Poiseuille flow profile (white arrows). With flow, the concentration in the center of the channel essentially equals that in the reservoir. Nonetheless, a significant diffuse boundary layer remains at the dissolving surface, with a boundary layer thickness of $\delta(0) \approx 0.5$ mm. The thickness of the boundary layer scales as $\delta \propto (x + H)^{1/3}$. The boundary layer, which is much smaller than the channel radius, reduces the effective advection rate by orders of magnitude, as the fluid velocity is low near the channel wall.

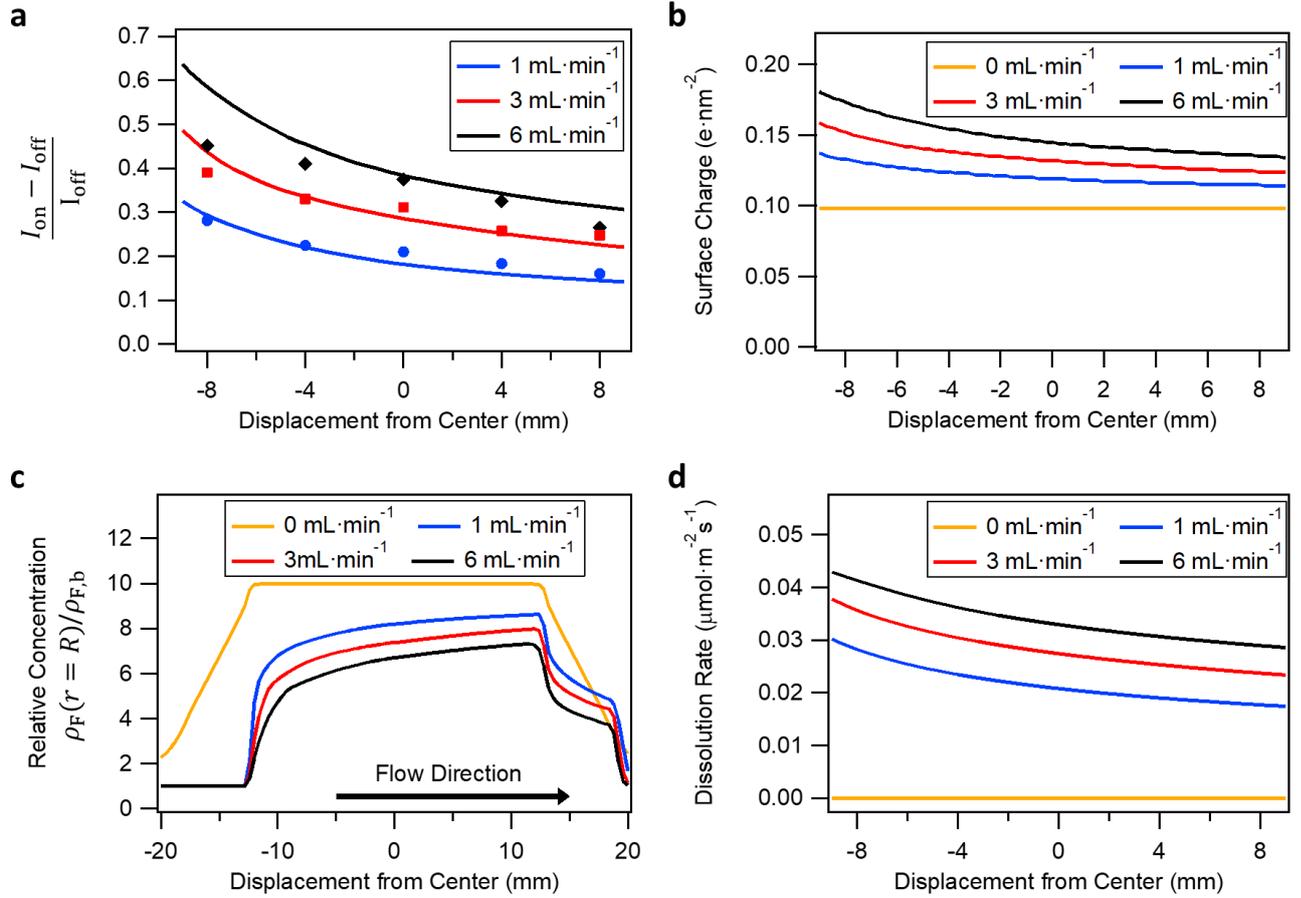

**Fig. 6. Results from the numerical model.** (a) Calculated (lines) and experimentally obtained (symbols) relative changes in the v-SFG intensity along the mineral surface for different flow rates. For clarity, we show only the averaged values from Fig. 3 without the spread of the individual data points. (b) Calculated surface charge along the mineral surface for different flow rates, showing enhancements under flow-on conditions by approximately 50% compared to flow-off conditions. (c) Calculated fluoride concentration at the surface $\rho_F(x, R)$ relative to the bulk along the mineral surface for several flow rates. (d) The calculated net dissolution rate $\mathbf{J}(x)$ along the mineral surface for different flow rates. Note the orders of magnitude increase upon turning on the flow. Under flow-off conditions, the dissolution rate is a small fraction of the maximum dissolution rate $\mathbf{J} \approx 10^{-5} k_{dis}$ while during flow the net dissolution increases by several orders of magnitude, to approximately half the maximal dissolution rate $\mathbf{J} \approx k_{dis}$. The position is given as displacement from the center with values increasing in the direction of flow, indicated by the arrow in panel (c).

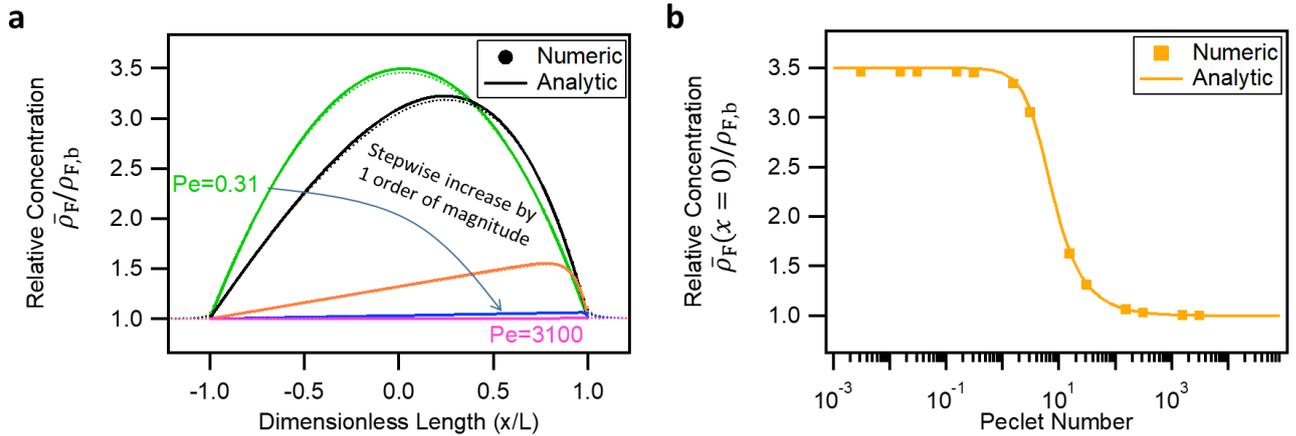

**Fig. 7. Numerical and Analytic results in the long channel limit.** (a) Numerical (dashed) and analytic (solid) results for the radially averaged fluoride concentration profile in the channel as a function of the distance from the center for Peclet numbers ranging from Pe = 0.31 (average flow velocity $3.2 \cdot 10^{-8}$ m·s$^{-1}$, green) to 3100 (average flow velocity $3.2 \cdot 10^{-4}$ m·s$^{-1}$, pink) in steps of a factor 10. (b) Radially averaged concentration at the center of the channel as a function of the Peclet number, comparing analytic (solid lines) and numerical (symbols) results for geologically realistic values ($k_{dis} = 10^{-11}$ mol·m$^{-2}$·s$^{-1}$, $L = 10^{-2}$ m, 10< average flow velocity <10$^{-4}$ m·s$^{-1}$, $D$ =10$^{-9}$ m$^2$·s$^{-1}$, $\eta$=10$^{-3}$ Pa·s). Note that the considered geometry has length $L = 1$ cm, radius $R = 0.1$ mm meeting the conditions of a large aspect ratio, which is not the case for the experimental geometry.

| Surface is \ Charging is | Desorptive | Adsorptive |
|---|---|---|
| Source | + | − |
| Sink | − | + |

**Fig. 8. Overview of Charging and Dissolution Coupling.** Indication whether the surface charge increases (+) or decreases (-) upon flow. Column position indicates if the surface is a source or sink of reactive ions, row position indicates if the reactive ion desorbs or adsorbs to generate surface charge. The change of surface charge upon flow is always larger at the inlet than at the outlet

*Supplementary Information on*

# Liquid Flow Reversibly Creates a Macroscopic Surface Charge Gradient

Patrick Ober[1], Willem Q. Boon[2], Marjolein Dijkstra[3], Ellen H. G. Backus[1,4], René van Roij[2*] & Mischa Bonn[1*]

[1]Department of Molecular Spectroscopy, Max Planck Institute for Polymer Research, Ackermannweg 10, 55128 Mainz, Germany. [2]Institute for Theoretical Physics, Utrecht University, Princetonplein 5, 3584 CC Utrecht, Netherlands. [3]Soft Condensed Matter, Debye Institute for Nanomaterials Science, Utrecht University, Princetonplein 5, 3584 CC Utrecht, Netherlands. [4]Department of Physical Chemistry, University of Vienna, Waehringer Strasse 42, 1090 Vienna, Austria.

These authors contributed equally: Patrick Ober, Willem Q. Boon

*email: r.vanroij@uu.nl, bonn@mpip-mainz.mpg.de

## Supplementary Note 1.

## Details on v-SFG and how it reports on the surface charge

Within the dipole approximation, incident light induces a dipole moment, which on a macroscopic scale (averaging over many molecules), is the polarization. For weak incident fields, the induced dipole moment ($\mu$) or polarization ($P$) scales linearly with the electric field. However, for laser pulses also higher-order terms need to be taken into account. For the second- and third-order terms, the proportionality constants between the applied field and the induced dipole are the first- and second-order hyperpolarizability $\beta$ and $\gamma$, respectively. On the macroscopic, polarization scale, this is the second and third-order nonlinear susceptibility $\chi^{(2)}$ and $\chi^{(3)}$[1]

$$\mu = \mu_0 + \alpha E + \beta E^2 + \gamma E^3 + \cdots$$
$$P = P^{(1)} + P^{(2)} + P^{(3)} + \cdots = \epsilon_0 \big(\chi^{(1)} E + \chi^{(2)} E^2 + \chi^{(3)} E^3 + \cdots \big). \quad (S1)$$

The total electric field is the sum of the electric fields of the two incident laser pulses

$$E = E_{\text{vis}} \cos(\omega_{\text{vis}} \cdot t) + E_{\text{IR}} \cos(\omega_{\text{IR}} \cdot t), \quad (S2)$$

Where the amplitudes are given by $E_{vis}$ and $E_{IR}$ and the frequencies by $\omega_{vis}$ and $\omega_{IR}$. From combining equations (S2) and (S3), a second-order term with the sum of the incident frequencies arises[1]

$$P_{\text{SFG}}^{(2)} = \epsilon_0 \chi^{(2)} E_{\text{vis}} E_{\text{IR}} \cos\big((\omega_{\text{vis}} + \omega_{\text{IR}}) \cdot t\big). \quad (S3)$$

Considering that the intensity of the sum-frequency light scales with the square of the polarization, we can derive the relation[1-3]

$$I_{\text{SFG}} \sim E_{\text{vis}}^2 E_{\text{IR}}^2 \big|\chi^{(2)}\big|^2, \quad (S4)$$

where $\chi^{(2)}$ is the macroscopic average of the first-order hyperpolarizability. In the case of a centrosymmetric system, $\chi^{(2)}$ becomes zero since the individual molecular hyperpolarizabilities cancel each other out, explaining the interface-specificity of v-SFG. At the interface, it is also crucial that a net-orientation of the molecules is present. In the case of random orientation, there is again canceling of the individual contributions. For the mineral-water interface, a net-orientation of interfacial water molecules is achieved by the presence of a surface charge.[1,2,4,5]

The presence of surface charge means that there is also an additional electric field present, which polarizes into the bulk and thereby breaks the symmetry requiring the addition of a $\chi^{(3)}$ contribution. The static, $DC$ electric field determines the magnitude of the $\chi^{(3)}$ contribution[5-8]

$$I_{\text{SFG}} \sim E_{\text{vis}}^2 E_{\text{IR}}^2 \left| \chi^{(2)} + \chi^{(3)} \int_0^\infty E_{\text{DC}}(z)\, dz \right|^2. \quad (S5)$$

We note that $\chi^{(2)}$ in Equation (S5) is effectively integrated over the interfacial layer where $\chi^{(2)}$ is non-zero. Within the Gouy-Chapman model, the electric field decays into the bulk solution as described by the Debye length. The Debye length is shorter with increasing salt concentration.[5-7]

Our measurements are performed in total internal reflection geometry, which is known to enhance the detected signal by orders of magnitude.[9] Therefore, the acquisition time of spectra can be reduced to seconds allowing the real time tracking of the dynamics when turning flow on and off.[4] Here we use this particular property to observe the equilibration of the system, e.g. converging to a steady state. The enhancement of the signal in total internal reflection geometry can be described by Fresnel factors.[9] One would have to take them into account when interfacial molecules are directly investigated by the spectral shape. As it is the magnitude of the signal, which correlates with the surface charge and is therefore of interest for our study, we integrate our spectra and do not focus on the spectral shape. Nevertheless, we note that Fig. 1b from the main text does not suggest significant changes of the spectral shape upon flow. For details on a relation between the spectral shape and Fresnel factors, we refer to other studies.[2,10-14]

Using total internal reflection geometry also requires consideration of the penetration depth of the evanescent fields that are generated at the interface. The interplay between the arising coherence length and the Debye length is complex.[5-7,15,16] If the penetration depth is shorter than the Debye length, the penetration depth would determine to which extent the $\chi^{(3)}$-term contributes to the signal. However, for our ionic strength in the millimolar range, the penetration depth of the evanescent field is much larger than the Debye length. Thus, we neglect the decay of the evanescent field and use the description of GONELLA et al.[6]

$$I_{\text{SFG}} \sim E_{\text{vis}}^2 E_{\text{IR}}^2 \left| \chi^{(2)} + \chi^{(3)} \psi \cdot f_3 \right|^2. \quad (S6)$$

Here $f_3$ describes a correction term for coherence, which for our ionic strength is ~1 and can, therefore, be neglected. The surface potential $\psi$ is described by the Graham equation:

$$\psi = 2\frac{k_B T}{e} \sinh^{-1}(2\pi\sigma\, \lambda_B \lambda_D) = 2\frac{k_B T}{e} \sinh^{-1}\left(\frac{e\sigma}{\sqrt{8 k_B T N_A \rho_s \epsilon_0 \epsilon_r}}\right), \quad (S7)$$

where $k_B$ is the Boltzmann constant, $T$ the temperature, $e$ the elementary charge, $\sigma$ the number of charged surface groups, $\lambda_B$ the Bjerrum length and $\lambda_D$ the Debye length, $N_A$ the Avogadro constant, $\rho_s$ the background concentration in mol/m$^3$, $\epsilon_0$ the electric permittivity of vacuum, and $\epsilon_r$ the relative permittivity of water.

As discussed in the main text, the flow-induced changes in ionic strength are much lower than the background ionic strength, which is why screening is excluded as a cause of the observed changes. Therefore, and in accordance with other studies[4,17], we correlate the changes in the v-SFG response in this work solely to changes in the surface charge. This change in surface charge can influence not only the surface potential but also $\chi^{(2)}$ and thus both terms in equation (S5) in a complex manner [18]. For simplicity, we employ the following approximation

$$I_{SFG} \sim \psi^2. \tag{S8}$$

## Supplementary Note 2.

## Additional v-SFG Results

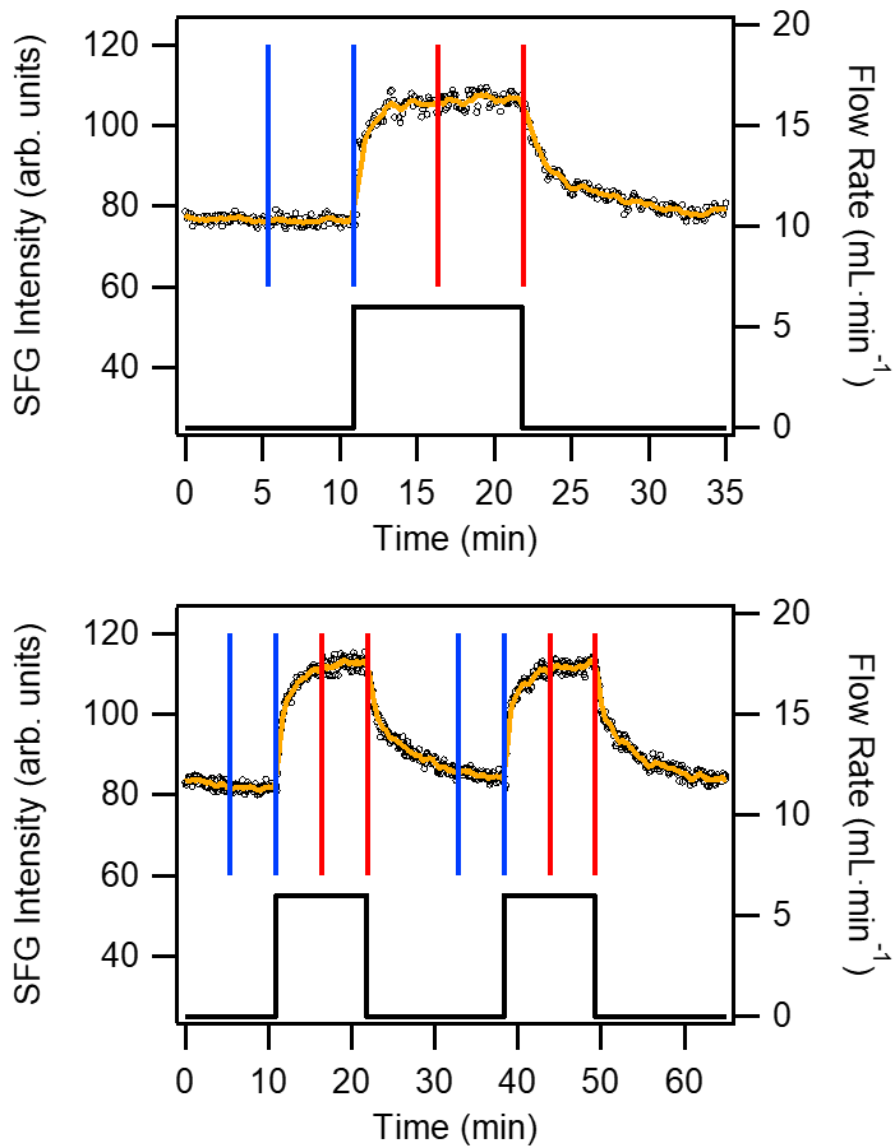

**Supplementary Figure 1. Exemplary time traces of the v-SFG signal with flow on/off cycles.** Similar to Fig. 1c in the main text, time traces of the integrated v-SFG signal (black circles) and a ten-point average (orange) as a guide for the eye are shown. The black curve shows the flow rate. The regimes marked by the blue and red lines are the flow-off and flow-on regimes, respectively. These regimes were used for calculating the relative increases in the v-SFG response due to flow. The change in the v-SFG intensity in these regimes is very low over time. Thus steady-state can be assumed.

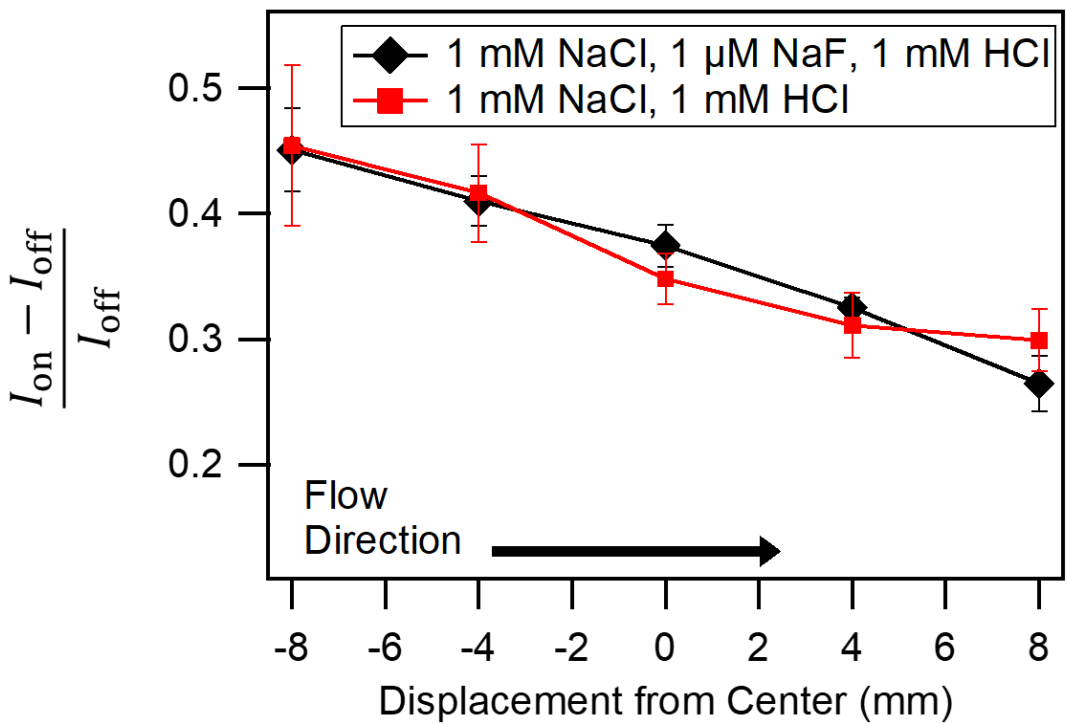

**Supplementary Figure 2. Gradient in the increase of the v-SFG intensity along the mineral surface.** At different positions, the relative increase in the v-SFG intensity at the $CaF_2$-water interface is calculated based on averaged steady-state spectra under flow-on and flow-off conditions as marked in Fig.1c of the main text and Supplementary Figure1. The open circles represent these ratios and the filled symbols are the mean of the ratios at the corresponding positions. The error of the individual ratios (open circles) due to small fluctuations within the steady-states of one flow-on/off cycle (see Fig1c of the main text and Supplementary Figure1) is negligible compared to the spread. The flow rate was 6mL/min. The solid lines are guides to the eye. The position is given as displacement from the center, with values increasing in the direction of flow, indicated by the arrow. The black data were already shown in Fig. 3 of the main text. By comparing the black and red data, it becomes clear that the addition of 1 µM NaF has no influence on the results as the points in black and red are within the spread of each other for every position.

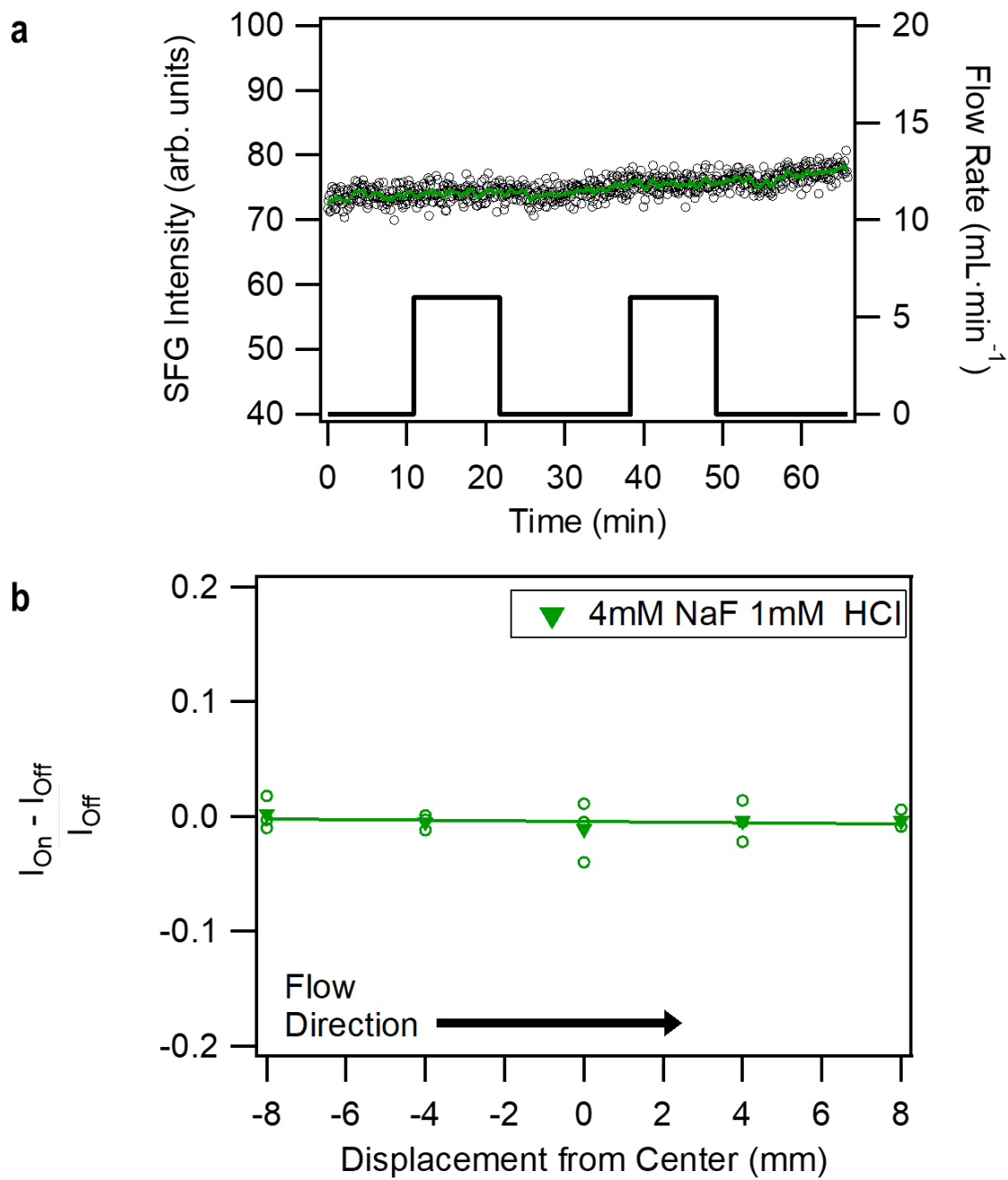

**Supplementary Figure 3. Experimental results at high fluoride concentrations.** (a) Exemplary time traces of the v-SFG signal with flow on/off cycles when using a 4 mM NaF solution in pH 3 HCl. (b) At different positions, the relative increase in the v-SFG intensity at the $CaF_2$-water interface is calculated based on averaged steady-state spectra under flow-on and flow-off conditions as marked in Fig1c of the main text and Supplementary Figure1. The open circles represent these ratios and the filled symbols are the mean of the ratios at the corresponding positions. The flow rate was 6mL/min. The solid line is a guide to the eye. The position is given as displacement from the center, with values increasing in the direction of flow, indicated by the arrow.

## Supplementary Note 3.

## Details of the Experimental Set up.

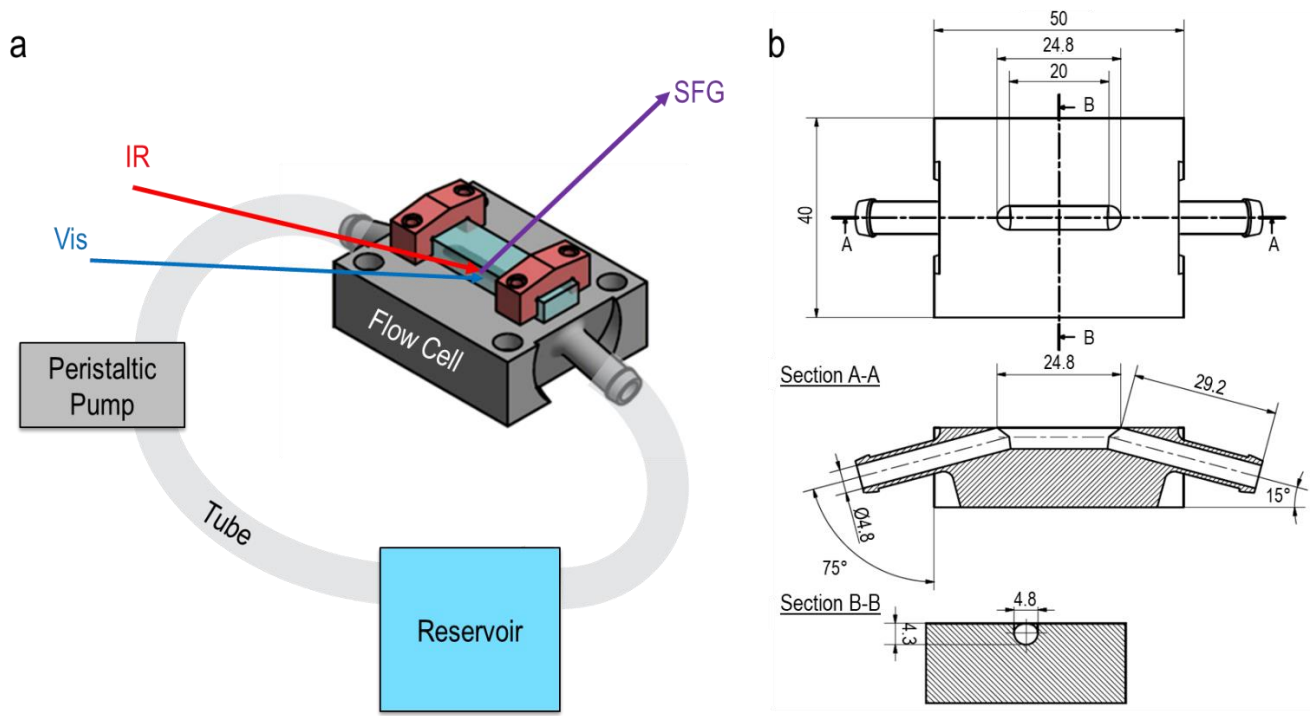

**Supplementary Figure 4. Illustration of Flow Set up.** (a) The flow cell has a flow channel with a rectangular $CaF_2$ prism mount on top. The Vis and IR laser pulse overlap at the center of the flow channel and generate SFG light in total internal reflection geometry. The flow cell is connected to a reservoir and a peristaltic pump. (b) Technical drawings of the flow cell with lengths in mm. Top panel shows a top view. A and B indicates the position of the cross sections in the middle and bottom panel, respectively.

## Supplementary Discussion 1.

## Full Poisson-Nernst-Planck-Stokes calculations

Solving the Poisson-Nernst-Planck equations with a finite element analysis requires a mesh size significantly smaller than the Debye length (~10 nm). We observed that the full experimental geometry has to be modeled, with a length scale of centimeters, to describe the experiments. The use of many more than $10^7$ mesh elements is computationally prohibitive. In this section, we will show that the results of small-scale Poisson-Nernst-Planck calculations are qualitatively similar and highlight a few minor quantitative deviations.

The set of equations that will be solved are the same as those in equations (3), (5)-(7) in the main text, but with the addition of the Poisson equation for the electric potential $\psi$, a conduction term added to the diffusion-advection equation, an extra electrostatic boundary condition relating the total surface charge density $e\sigma$ to the electric field over the surface normal $\mathbf{n}$. For completeness, a body force term $e \sum_i z_i \rho_i \, \nabla\psi$, with $z_i$ the ion valency, has to be added to the Stokes equation. However, this term can be almost always neglected in pressure-driven flows. The set of equations is thus given by

$$\nabla^2 \psi = -\frac{e}{\epsilon} \sum_i z_i \rho_i, \tag{S9}$$

$$\partial_t \mathbf{u} = -\nabla p + \eta \nabla^2 \mathbf{u} - e \sum_i z_i \rho_i \, \nabla\psi, \tag{S10}$$

$$\partial_t \rho_i = D \nabla^2 \rho_i - \nabla \cdot (\rho_i \mathbf{u}) + D \frac{e z_i \rho_i}{k_B T} \nabla\psi, \tag{S11}$$

$$\mathbf{n} \cdot \nabla\psi = \frac{e\sigma}{\epsilon}, \tag{S12}$$

$$\psi(\pm\infty) = 0. \tag{S13}$$

In equilibrium this set of equations reverts back to regular Poisson-Boltzmann theory, that yields a diffuse layer of excess counter-ions, the electric double layer, near the charged surface. The decay length of the electric potential in this layer is the Debye length $\lambda_D = (8\pi\lambda_B\rho_s)^{-1/2}$ where $\lambda_B$ is the Bjerrum length and $\rho_s$ the bulk concentration of an added 1:1 electrolyte. The geometry is scaled down to a cylindrical channel with length 5 μm and radius 0.15 μm. The length of the dissolving, charged surface in the middle of the channel is 2.5 μm. When including dissolution, a concentration profile of reactive ions forms, which breaks the translational invariance along the channel. When the concentration gradient over the Debye length is much larger than the concentration gradient over the channel, $\partial_r \rho \gg \partial_x \rho$, we can use the lubrication layer approximation[19]. This allows us to write the surface charge as $\sigma(x) = \Gamma\big(1 + k_{ads}\rho(r = 0, x) \, e^{\phi_0(\sigma(x))}/k_{des}\big)^{-1}$, with $\phi_0 = \psi_0 \, e/k_B T$ the dimensionless electric potential at the surface. We combine this result with the Gouy-Chapman result derived for the surface potential of flat plates finding $\phi(x) = 2 \sinh^{-1}(2\pi\lambda_B\lambda_D\sigma(x))$. We compare the Gouy-Chapman surface potential with the numerically calculated surface potential in Supplementary Figure 5. Note that both the lubrication layer and Gouy-Chapman approximation work well at the $\mu m$ length scale, and the quality of the approximations is expected to increase at larger length scales.

When a pressure difference is applied over the channel, the electric boundary condition becomes of importance, as a streaming current is generated. In the experimental system, the fluid circulates, and inlet and outlet are connected, so boundary conditions representing a closed electric circuit are used: $\psi(x = \pm\infty) = 0$. The resulting concentration profiles for Peclet numbers of one, two, and four times $3.5 \cdot 10^6$ can be seen in Supplementary Figure 5. These calculations were repeated without any charge on the dissolving surface, but with the same surface reaction. The concentration profile in the middle stays the same with and without the conduction term.

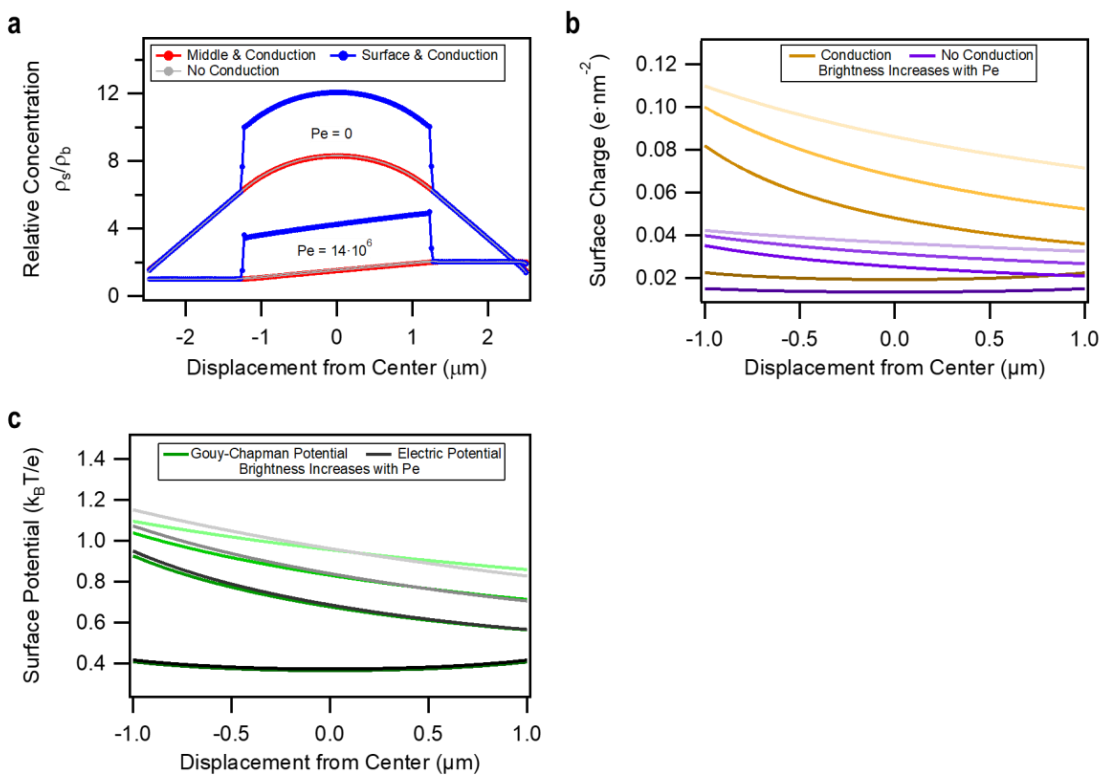

**Supplementary Figure 5. Comparison between Small Scale Calculations with and without a Conduction Term.** (a) The concentration on the central axis of the channel (red) and the concentration at the surface (blue) with a conduction term for the lowest and highest Peclet number investigated here, together with the corresponding profiles (gray) without conduction, which both overlap almost exactly with the red curves. The difference between the red and blue lines is due to electrostatic attraction between the surface and reactive ions, and can be accounted for by a Boltzmann factor $e^{\phi_0}$. (b) surface charge heterogeneity with conduction (orange) and without conduction (violet) for Peclet numbers 0, 1, 2 and 4 times $3.5 \cdot 10^6$ (dark to light). The large quantitative difference is again due to a change in concentration due to electrostatic attraction, but qualitatively the trends are the same. (c) Comparison between the electrostatic potential directly extracted from the numerical calculations with conduction (gray) and the surface potential as calculated by the Gouy-Chapman (green) from the same calculation for Peclet numbers 0, 1, 2 and 4 times $3.5 \cdot 10^6$ (dark to light).

The chief difference between these two concentration profiles is an increase of the concentration by a factor of $e^{\phi_0}$ at the surface. While this does not change any of the qualitative features of our model, namely a surface charge gradient induced by flow over a dissolving surface, it does affect the magnitude of this effect. This can be seen in Supplementary Figure 5b where we plot the obtained surface charge with and without conduction at the different flow rates. The difference between the two responses is significant (approximately a factor of five). This is mainly due to a rescaling of the equilibrium constant $K$ by a Boltzmann factor $K_{ch} = K_{unch}e^{\phi_0}$, where $K_{ch}$ is the equilibrium constant $k_{ads}\rho_b/k_{des}$ with electrostatics and $K_{unch}$ is the equilibrium constant without electrostatics. As the equilibrium constant is an effective fit parameter used to reproduce the equilibrium surface potential $\psi_0$, this rescaling is effectively absorbed during our calculation in the main text. Note that the equilibrium surface potential is calculated using $K_{ch}$ instead of $K_{unch}$, which therefore accounts for the electric double layer. When the surface potential changes, the equilibrium constant also changes. This feature is neglected in the main text, however it only introduces minor quantitative deviations. Thus, the main conclusions from the numerical calculations with electrostatics included are that (i) the expression for equilibrated EDL's (Gouy-Chapman) is valid and hence non-equilibrium electrostatic phenomena can be neglected, and (ii) upon including electrostatics the same qualitative behavior is found as in the main text.

## Supplementary Discussion 2.

### Analytic one-dimensional model

As analytically solving the two-dimensional dissolution-diffusion-flow problem defined by equations (6)-(7) in the main text is intractable, we first convert the problem to an effective one-dimensional problem, which is analytically solvable. This one-dimensional problem can be solved while taking most experimental complications into account (back-reaction, a piece of the channel without a dissolving surface). However, the resulting analytic results are usually too complex to aid understanding of the system. Therefore, we introduce some further simplifications. The presented method can be straightforwardly extended to a parallel plate geometry.

We transform the two-dimensional diffusion-advection problem to a one-dimensional problem by radially integrating the diffusion-advection equation to obtain the average density $\bar{\rho}(x) = 2\pi \int \rho(x,r)\, r\, dr/\pi R^2$. For convenience we drop the subscript F for the entirety of this section, but will still focus on the ion partaking both in charging and dissolution i.e. the fluoride concentration. In the regions without dissolution, we then find

$$\partial_t \bar{\rho} = D\partial_x^2 \bar{\rho} - \partial_x \overline{u\rho}. \tag{S14}$$

Whereas the diffusive term is straightforwardly integrated, the calculation of the advective term is more complicated. This is because it results in a density-weighted velocity $\overline{u\rho} = 2\pi \int u_x(r)\rho(r,x)\, r\, dr/\pi R^2$, which requires the knowledge of $\rho(x,r)$ as a function of the radial coordinate to calculate. In the narrow-channel limit $\frac{\delta}{R} \gg 1$, however, the approximation $\rho(x,r) = \bar{\rho}(x)$ is valid as shown in the main text, and the radially averaged advection-diffusion equation simply becomes:

$$\partial_t \bar{\rho} = D\partial_x^2 \bar{\rho} - \bar{u}\partial_x \bar{\rho}. \tag{S15}$$

In the opposite limit of a thin boundary layer ($\frac{\delta}{R} \ll 1$) the integral $\overline{u\rho}$ does not factorize as there is a significant radial variation in the concentration $\rho(x,r)$. A qualitative approximation can be made in the limit of high Peclet (Pe $\gg$ 1) by replacing the average velocity $\bar{u}$ by the effective boundary transfer velocity $h \propto \bar{u}^{2/3}$, as will be derived in Supplementary Discussion 2. The boundary transfer velocity accounts for the effect that at high flow velocities the ions released from the surface do not have time to spread out over the length of the entire radius by diffusion. This lowers the effective flow velocity ($h \propto \bar{u}^{2/3}$) as the velocity near the surface of the channel is lower than at the center.

Continuing with the narrow channel limit $\frac{\delta}{R} \gg 1$, where $\overline{u\rho} = \bar{u}\bar{\rho}$, the integration of equation (S15) in the region with a dissolving surface (represented by the source term **J**) is straightforward and results in

$$\partial_t \bar{\rho} \approx D\partial_x^2 \bar{\rho} - \bar{u}\partial_x \bar{\rho} + 2n\frac{k_{\text{dis}} - k_{\text{prec}}\rho_b}{R} - 4n\frac{k_{\text{prec}}}{R}(\bar{\rho} - \rho_{\text{b}}), \quad (S16)$$

where the dissolution term was substituted for the radial concentration gradient at the boundary $\mathbf{n} \cdot \mathbf{J} = D\partial_r \rho$. We linearized the precipitation in the density $k_{\text{prec}}\rho^n \approx k_{\text{prec}}\rho_{\text{b}} - n\,k_{\text{prec}}(\rho - \rho_{\text{b}})$ to ensure that the equation remains a linear, solvable differential equation. Here $n$ is the stochiometric constant relating the number of ions released per uncharged unit. Note that these equations are approximations valid in the limit $\frac{\delta}{R} \ll 1$.

From differential equations equation (S15) and equation (S16) we can construct analytic solutions for simple geometries involving patches of non-dissolving and dissolving channels. For instance, in a channel with a dissolving surface in the middle separated by two regions without dissolution to two bulk reservoirs, we would have three differential equations, namely two patches without dissolution and one patch with dissolution. For the dissolving surface, we would use equation (S16), and for the regions without a dissolving surface, equation (S15) is used. At the edge of each discontinuity, the boundary condition $\bar{\rho}_n(x_n) = \rho_n$ is used, where $x_n$ is the boundary between geometry elements $n$ and $n+1$, $\bar{\rho}_n$ denotes the solution to the $n$th differential equation. The boundary concentration $\rho_n$ is an unknown concentration that can be fixed by using that the average concentration $\bar{\rho}_n$ at each boundary is continuous $\bar{\rho}_n(x_n) = \bar{\rho}_{n+1}(x_n)$, which is true as long as the radius $R(x)$ is continuous. This then gives $n-1$ equalities for $n+1$ unknown concentrations. The two unknown concentrations are located at the edges of the model system (for instance, the edge of the bulk reservoirs). In this work, the boundary condition $\bar{\rho}_1(x_0) = \bar{\rho}_N(x_N) = \rho_b$, with $x_0$ being the leftmost system boundary and $x_N$ being the rightmost system boundary, was always used. While this set of equations is always analytically solvable (and is straightforwardly extended to, for instance, multiple reactive patches) the resulting expressions are often so complicated to the point of being illegible, even with only three spatial regions (two non-dissolving patches and a channel with only a dissolving surface). The solutions are computationally accessible, so the solutions can be straightforwardly plotted to gain insight without having to resort to numerical calculations. Here we will consider the solution in a simplified case, leading to the result shown in the main text. We consider a geometry with only one part; a dissolving channel of length $2L$, with radius $R$, and neglect the precipitation term. We chose not to add the two non-dissolving walls because of the additional complexity in the final expressions. The resulting expression is

$$\frac{\rho(x)}{\rho_{\text{b}}} = 1 + \frac{\Delta \rho_{\max}}{\text{Pe}}\left(\frac{2x}{L} + \left(1 + e^{2\,\text{Pe}} - 2e^{\text{Pe}\left(1+\frac{x}{L}\right)}\right)(\text{Coth}(\text{Pe}) - 1)\right), \quad (S17)$$

which, in the limit of low and high Peclet number, simplifies to equation (8) in the main text. Here $\Delta\rho = k_{\text{dis}}L^2/DR\rho_{\text{b}}$ and $\text{Pe} = 2uL/D$. It can be seen in Figure 7 of the main text that this analytic result matches numerical results in the narrow channel limit ($\delta/R \gg 1$).

In principle, this one-dimensional model is fully solvable for any radially symmetric geometry with continuous radius $R(x)$ and constant radial concentration $\rho(x, r) = \bar{\rho}(x)$. In practice, the resulting expressions, while valid, become so complex that they are useless for insight and even become computationally heavy to plot and manipulate. Highly simplified analytic models lose quantitative predictive power but incorporate all the fundamental physics in the problem: dissolution, advection, and reactions, and qualitatively describes how they balance each other.

## Supplementary Discussion 3.

### Diffuse boundary layer

In this section we summarize the derivation of the diffuse boundary layer as presented elsewhere[20]. A diffuse boundary layer is the concentration profile that forms near a dissolving surface when lateral advection is so strong that there is not enough time for radial diffusion to occur. As in the previous section, we drop the subscript $F$ on the density $\rho$, but we are still explicitly solving for the fluoride concentration.

Observing that at high Peclet numbers lateral diffusion can be neglected, the diffusion-advection equation can be written in stationary state as

$$\partial_t \rho = D\nabla^2 \rho - \nabla \cdot (\mathrm{u}\rho) \approx \frac{D}{r} \partial_r (r \partial_r \rho) - u_x(r) \partial_x \rho = 0. \tag{S18}$$

Now we consider a very long channel with a fully developed laminar flow profile with a laterally *constant* surface concentration $\rho_\mathrm{m}$, for which the boundary conditions can be written as

$$\begin{cases} \rho(\pm\infty, r) = \rho_\mathrm{b}, \\ \rho(x, R) = \rho_\mathrm{m}, \\ \partial_r \rho(x, 0) = 0. \end{cases} \tag{S19}$$

This is traditionally known as the Graetz problem[20-22]. The assumption of a constant surface concentration is not valid in the model considered in the main text. However, the present analysis is concerned with describing the concentration profile some distance $\delta$ from the surface. We will solve differential equation (S18) together with boundary conditions equation (S19) using the Ansatz that the solution is self-similar, and that the concentration profile $\rho(x, r)$ that depends on two variables can be described by a function of a single self-similar variable $\eta$

$$\rho(x, r) = f(\eta). \tag{S20}$$

The existence of a self-similar solution is suggested[23] because it is not possible to define a dimensionless lateral position $X$ using only the lateral position and velocity, $x$ and $u_x$, and hence the concentration should be reducible to a one parameter equation of $\eta \propto (R - r)/\delta(x)$. Now we define the dimensionless, self-similar, parameters

$$\eta = \frac{R-r}{\delta(x)} = (R-r)\left(\frac{4\bar{u}}{9DR(x+H)}\right)^{\frac{1}{3}}, \tag{S21}$$

$$\zeta = \frac{\delta(x)^3}{R^3} = \frac{9D(x+H)}{4\bar{u}R^2}, \tag{S22}$$

$$\Theta = \frac{\rho - \rho_m}{\rho_b - \rho_m}. \tag{S23}$$

and using this we rewrite equation (S19) to[20]

$$\frac{\partial^2 \Theta}{\partial \eta^2} + 3\eta^2 \frac{\partial \Theta}{\partial \eta} - 3\eta\zeta^{\frac{1}{3}} \frac{\partial \Theta}{\partial \zeta^{\frac{1}{3}}} = \left(\frac{3}{2}\eta^3\zeta^{\frac{1}{3}} + \frac{\zeta^{\frac{1}{3}}}{1-\eta^3\zeta^{\frac{1}{3}}}\right)\frac{\partial \Theta}{\partial \eta} - \frac{3}{2}\eta^2\zeta^{\frac{2}{3}}\frac{\partial \Theta}{\partial \zeta^{\frac{1}{3}}}. \tag{S24}$$

We find the dimensionless boundary conditions

$$\begin{cases} \Theta(\eta = \infty) = 1, \\ \Theta(\eta = 0) = 0. \end{cases} \tag{S25}$$

The solution to equation (S24) can be expanded as function of $\zeta = \delta/R$

$$\Theta = \Theta_0(\eta) + \zeta^{\frac{1}{3}}\Theta_1(\eta) + \zeta^{\frac{2}{3}}\Theta_2(\eta) + O(\zeta), \tag{S26}$$

where at high Peclet number we only consider the dominant zero-order term, also known as the Lévêque approximation

$$\frac{d^2\Theta_0}{d\eta^2} + 3\eta^2 \frac{d\Theta_0}{d\eta} = 0. \tag{S27}$$

This is approximation is equal to only considering the first order term of the velocity expansion $u(r) \approx 4\bar{u}(1 - r/R) + O((1 - r/R)^2)$ [21] close to $r = R$, and hence is valid in the limit of high Peclet number where $\delta/R \ll 1$. Solving for $\Theta_0$ we find the Lévêque solution

$$\Theta_0(\eta) = C^{-1} \int_0^\eta e^{-s^3} \, ds, \tag{S28}$$

$$C = \int_0^\infty e^{-s^3} \, ds = \Gamma\left(\frac{4}{3}\right) \approx 0.9. \tag{S29}$$

with $\Gamma$ the Euler gamma function. Plotting the concentration profile in Supplementary Figure 6 as a function of the radial coordinate at $x = 0$ we see that the radial concentration profile is very well approximated by the Lévêque solution. In Supplementary Figure 7 we plot the radial position $R - r$ at which the concentration satisfies $1.9 < \frac{\rho_F(r)}{\rho_{b,F}} < 2.1$ together with $\delta(x)$ as function of $(x + H)^{1/3}$. We find that the boundary layer thickness in our numerical calculations is linear in $(x + H)^{1/3}$ as expected.

Thus, for $x = 0$ and at a flow rate of 1 mL/min we find that the typical boundary layer thickness at which the radial concentration has decayed by 90% of its surface value is approximately $\delta \approx 0.4$ mm.

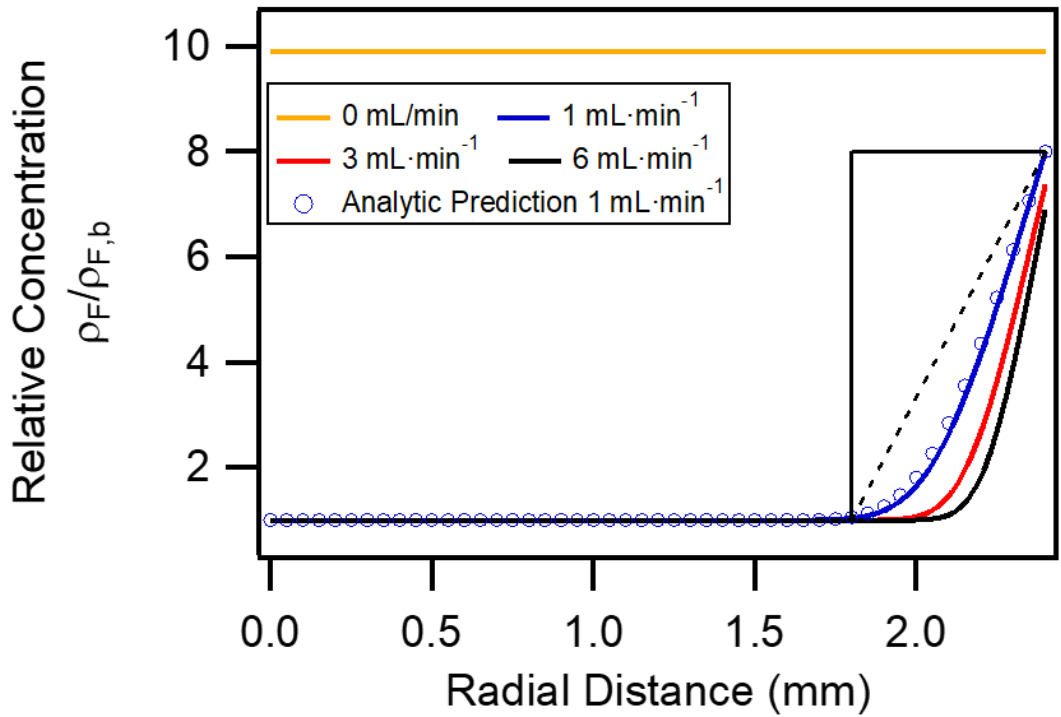

**Supplementary Figure 6. Radial Concentration Profiles.** The radial concentration profile of reactive ions in the center of the channel at different flow speeds are shown. The data were obtained from full numerical calculations as described in the main text. It can be seen that close to the surface, the concentration profile is well described by a linear concentration profile with a flow-dependent length scale. Circles are the analytically predicted diffuse boundary layer at the lowest flow rate according to equation (S28) with the maximum density $\rho_m$ fit parameter.

While the behavior of the concentration profile at some distance from the channel wall is thus correctly described by diffuse boundary layer theory, the near-surface $(R - r \ll \delta)$ profile deviates significantly from the Lévêque solution as the assumption of a constant surface concentration is incompatible with the macroscopic concentration gradient, which is the main object of interest in the main text. Thus, while superficially similar to a cuberoot scaling, the surface concentration $\rho(x, R)$ does not follow the power law $(x + H)^{1/3}$. To estimate the influence of the diffuse boundary layer on the surface concentration $\rho(x, R)$ we insert an approximation for the radial concentration profile (S34, S36) into our analytic calculation as presented in Supplementary Discussion 2, which allows us to calculate $\rho_m = \rho(x, R)$. Note that this combination of diffuse boundary layer theory and the radially integrated diffusion-advection equation does not result in a fully self-consistent theory. However it does allow us to estimate the change in surface concentration with regard to flow-rate.

The fact that the diffuse boundary layer is localized near the channel wall where the velocity is low implies that the effective velocity for the transport of ions is much lower than the channel-averaged velocity. To estimate the effect of the small boundary layer thickness we need an expression for $\rho(x, r)$ such that the integral $\int_0^R \rho(r) u(r) r \, dr$ can be evaluated, as required for S14. Inspecting the numerically obtained concentration profiles in Supplementary Figure 5 we see that the it can be approximated by the simple form

$$\rho(r, x) = \begin{cases} \rho_b & \text{if } r < R - \delta(x), \\ \rho_b + \dfrac{\rho_m - \rho_b}{\delta(x)}(r - R + \delta(x)) & \text{if } R - \delta(x) < r < R. \end{cases} \quad \text{(S30)}$$

With this approximation the divergence of the density weighted velocity $\overline{u\rho}$, as defined in equation (S14), can be straightforwardly found

$$\partial_x \overline{u\rho} = \partial_x \left( \bar{u} \frac{4\delta^2}{3R^2} (\rho_m - \rho_b) \right) + O\left(\frac{\delta^3}{R^3}\right) \approx \partial_x \frac{4\bar{u}\delta}{3R} \bar{\rho}. \quad \text{(S31)}$$

From this we can define the effective density weighted velocity $h$, also known as the boundary transfer velocity[22]. The boundary transfer velocity is given by $h = 4\bar{u}\delta/3R$ when the surface concentration is much larger than the bulk concentration ($\rho_m \gg \rho_b$), in which case $\bar{\rho} = \rho_m \delta/R$, and the gradient in the surface concentration is larger than the gradient in the boundary thickness ($\rho_m^{-1}\partial_x\rho_m \gg \delta^{-1}\partial_x\delta$). As seen from S28 the boundary thickness satisfies $\delta \propto \bar{u}^{-1/3}$ and thus the effective transfer velocity scales as

$$h = \bar{u}\frac{4\delta}{3R} \propto \bar{u}^{2/3}. \quad (S32)$$

From this it can be straightforwardly seen that the Sherwood number Sh, which is the dimensionless number comparing the boundary transfer rate with the diffusion rate, scales as Sh= $2Lh/D \propto \text{Pe}^{2/3}$ [20,22]. While the effective advection rate scaling by a power of 1/3 with the channel-averaged advection rate is a well-known result for a channel wall, kept at a *constant* concentration[20-22], here we show that a phenomenological model for a dissolving channel with a *heterogeneous* surface concentration gives rise to a power of 2/3.

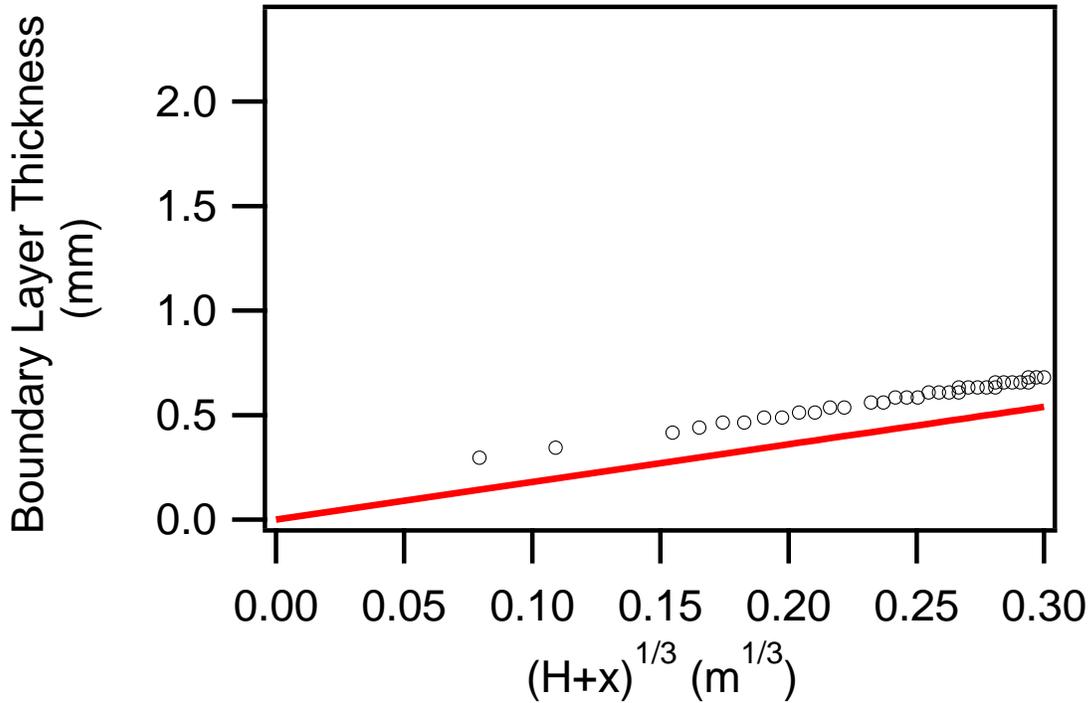

**Supplementary Figure 7. Verification cube root scaling diffuse boundary layer thickness.** The radial position at which the concentration is between $1.9 < \rho/\rho_b < 2.1$ (data points) against the cube root of the lateral position $(H+x)^{1/3}$, plotted from the $x = -H$ to $x = H$. It can be clearly seen that the numerical data is linear in the cube root of the position $(H+x)^{1/3}$. The y-axis scale is between $0 < r < R$.

The Sherwood number replacing the Peclet number and scaling as a power smaller than 1 of the latter captures an important aspect of our experiment, namely that the effective advection rate over the boundary layer is much lower than the channel-averaged advection rate. This causes the sharp transition between the flow and no-flow concentrations seen in Supplementary Figure 7b to broaden significantly to larger Peclet number. As the experimental Peclet number is Pe $\approx 10^5$, the Sherwood number is lower by almost two orders of magnitude Sh $\approx 10^3$. Combined with the precipitation reaction, which increases the dissolution rate at lower surface concentration, the lower effective advective transfer rate $h$ explains why there is still a significant change of concentration with flow while the limit of large Peclet number has been reached in our experiments. We stress that the scaling relation between surface concentration and flow rate in the case of a fully developed diffuse boundary layer is an order-of-magnitude estimation and that we have not found a regime in which it is quantitatively accurate.

**Supplementary References**